\documentclass[10pt]{iopart}
\usepackage{graphicx,epsfig}
\usepackage{bm}
\usepackage{iopams}
\newcommand{\ba}{\begin{eqnarray}}
\newcommand{\ea}{\end{eqnarray}}
\newcommand{\bse}{\numparts}
\newcommand{\ese}{\endnumparts}
\newcommand{\ACal}{{\cal{A}}}

\newcommand{\DD}{{\cal {D}}}

\newcommand{\bbq}{\begin{quote}}
\newcommand{\eeq}{\end{quote}}
\newcommand{\tbb}{t_{\textrm{\tiny{bb}}}}

\newcommand{\tbbo}{t_{\textrm{\tiny{bb(0)}}}}
\newcommand{\tcoll}{t_{\textrm{\tiny{coll}}}}

\newcommand{\tmax}{t_{\textrm{\tiny{max}}}}

\newcommand{\amax}{a_{\textrm{\tiny{max}}}}

\newcommand{\RR}{{\cal{R}}^{(3)}}
\newcommand{\RRR}{{\cal{R}}}

\newcommand{\FF}{{\cal{F}}}
\newcommand{\JJ}{{\cal{J}}}
\newcommand{\VV}{{\cal{V}}}
\newcommand{\HH}{{\cal{H}}}
\newcommand{\KK}{{\cal{K}}}

\newcommand{\Da}{\delta^{(A)}}

\newcommand{\Dh}{\delta^{(\HH)}}

\newcommand{\Drho}{\delta^{(\rho)}}

\newcommand{\DKK}{\delta^{(\KK)}}
\newcommand{\DOm}{\delta^{(\Omega)}}
\newcommand{\dd}{{\rm{d}}}

\newcommand{\Dig}{\Delta_0^{\tiny{\textrm{(g)}}}}
\newcommand{\Did}{\Delta_0^{\tiny{\textrm{(d)}}}}

\newcommand{\Jg}{\JJ_{\tiny{\textrm{(g)}}}}
\newcommand{\Jd}{\JJ_{\tiny{\textrm{(d)}}}}
\newcommand{\Aav}{\langle A\rangle}
\newcommand{\rhoav}{\langle \rho\rangle}
\newcommand{\HHav}{\langle \HH\rangle}
\newcommand{\Tgr}{T_{\tiny{\textrm{gr}}}}
\newcommand{\sgr}{s_{\tiny{\textrm{gr}}}}
\newcommand{\pgr}{p_{\tiny{\textrm{gr}}}}
\newcommand{\qgr}{q^a_{\tiny{\textrm{gr}}}}
\newcommand{\Pigr}{\Pi^{ab}_{\tiny{\textrm{gr}}}}
\newcommand{\rhogr}{\rho_{\tiny{\textrm{gr}}}}
\newcommand{\Del}{{\textrm{\bf{D}}}}
\newcommand{\Denl}{{\textrm{\bf{D}}}^{(\tiny{\textrm{NL}})}}
\newcommand{\shb}{s_{\textrm{\tiny{HB}}}}

\usepackage{pdfsync}
\begin{document}

\title[Gravitational entropies in LTB dust models]{Gravitational entropies in LTB dust models}
\author{Roberto A. Sussman$^{1}$ and Julien Larena$^{2}$}
\address{$^{1}$ Instituto de Ciencias Nucleares, Universidad Nacional Aut\'onoma de M\'exico (ICN-UNAM),
A. P. 70--543, 04510 M\'exico D. F., M\'exico.\\
$^{2}$ Department of Mathematics, Rhodes University, Grahamstown 6140, South Africa
}

\eads{$^{1}$\mailto{sussman@nucleares.unam.mx}, $^{2}$\mailto{j.larena@ru.ac.za}}

\date{\today}
\begin{abstract} We consider generic Lema\^\i tre--Tolman--Bondi (LTB) dust models to probe the gravitational entropy proposals of Clifton, Ellis and Tavakol (CET) and of Hosoya and Buchert (HB). We also consider a variant of the HB proposal based on a suitable quasi--local scalar weighted average. We show that the conditions for entropy growth for all proposals are directly related to a negative correlation of similar fluctuations of the energy density and Hubble scalar. While this correlation is evaluated locally for the CET proposal, it must be evaluated in a non--local domain dependent manner for the two HB proposals. By looking at the fulfilment of these conditions at the relevant asymptotic limits we are able to provide a well grounded qualitative description of the full time evolution and radial asymptotic scaling of the three entropies in generic models. The following rigorous analytic results are obtained for the three proposals: (i) entropy grows when the density growing mode is dominant, (ii) all ever-expanding hyperbolic models reach a stable terminal equilibrium characterized by an inhomogeneous entropy maximum in their late time evolution; (iii) regions with decaying modes and collapsing elliptic models exhibit unstable equilibria associated with an entropy minimum (iv) near singularities the CET entropy diverges while the HB entropies converge;  (v) the CET entropy converges for all models in the radial asymptotic range, whereas the HB entropies only converge for models asymptotic to a FLRW background. The fact that different independent proposals yield fairly similar conditions for entropy production, time evolution and radial scaling in generic LTB models seems to suggest that their common  notion of a ``gravitational entropy'' may be a theoretically robust concept applicable to more general spacetimes.                                                          
\end{abstract}

\maketitle
\section{Introduction.}

The notion of a self--consistent ``gravitational entropy'', distinct from (though possibly related with) the entropy of the sources (thermal sources or black holes) is an open  problem with interesting theoretical ramifications in General Relativity. This notion comes originally from Penrose's old idea \cite{arrow1} of the ``arrow of time'', associated with the ratio of scalars contractions of the Weyl and Ricci tensors. This idea was further developed and modified by different authors~\cite{arrow1,arrow2,arrow3,arrow4,arrow5,arrow6}. A more recent approach by Clifton, Ellis and Tavakol (CET) \cite{CET} no longer relies on invariant curvature scalars, but on an ``effective'' energy--momentum tensor associated with the ``free gravitational field'' and obtained from the Bell--Robinson tensor. An alternative approach is based on an entropy functional from the Kullback-Leibler divergence of Information Theory \cite{Info}, applied by Hosoya and Buchert (HB) \cite{HB1,HB2,HB3} to a cosmological context, and in a modified form (the ``HBq'' proposal denoting the original HB proposal by ``HBp'') to spherically symmetric Lema\^{\i}tre--Tolman--Bondi (LTB) dust models in \cite{part1} by means of the quasi--local weighted average instead of Buchert's average. 

In the present article we use generic LTB models to probe the CET, HBp and HBq gravitational entropy proposals, as a first step to test the theoretical solidity of their predictions and properties. While LTB models \cite{LTB} are highly idealized toy models, they are particularly well suited to understand and study a host of non--linear non--perturbative relativistic effects of cosmological and astrophysical self--gravitating systems by means of mathematically tractable methods (see the comprehensive reviews in \cite{kras1,kras2,BKHC2009,celerier,focus}). They have been used to describe a wide variety of phenomena: structure formation and late time cosmological inhomogeneities \cite{KH,ltbstuff}, fitting cosmological observations without resorting to dark energy (see \cite{kras2,BKHC2009,marranot} for a review), testing averaging formalisms \cite{LTBave1,LTBave2,sussBR,sussIU,suss2011}, cosmic censorship \cite{lemos,joshi} and even describing some effects in quantum gravity \cite{quantum}.

In order to use LTB models to examine and compare the gravitational entropy proposals, we describe their dynamics in terms of an initial value formulation based on an alternative representation of coordinate independent scalars (the ``quasi--local'' or ``q--scalars'') that follow from a weighted proper volume average on comoving domains (see \cite{part1,part2} for a comprehensive study).  These scalar variables have been very useful to look at the models under a dynamical systems approach \cite{sussDS1,sussDS2}, to examine their asymptotic behaviour in the radial direction \cite{RadAs}, the evolution of radial profiles and void formation \cite{RadProfs}, the existence of back--reaction and ``effective'' acceleration in the context of Buchert's formalism \cite{sussBR,sussIU,suss2011}, and even to study dark energy sources compatible with the LTB metric \cite{sussQL,suss2009}. As shown in \cite{part2}, the q--scalars and their fluctuations and perturbations lead to a covariant and gauge invariant formalism of ``exact perturbations'' on a Friedmann-Lema\^itre-Robertson-Walker (FLRW) abstract background defined by the q--scalars themselves (which satisfy FLRW time evolution laws). Hence, we find it  useful to express these exact perturbations in terms of an exact covariant generalization of the growing and decaying density modes of linear perturbations of dust sources (see comprehensive discussion in \cite{sussmodes}). We must also mention the dynamical studies of LTB models (with zero and nonzero $\Lambda$) by Wainwright and Andrews \cite{WainAndr} in terms of growing and decaying density modes (see the review of this article in Appendix B of \cite{sussmodes}) and in terms of frame variables by Coley {\it et al} \cite{coley}.  

It is known that the condition for positive entropy production for the HBp and HBq proposals is a negative statistical correlation between the fluctuations of the energy density and Hubble expansion scalar (see \cite{HB1,HB2,HB3,part1}).  We prove in this paper that the entropy production condition for the CET proposal is also a negative correlation between density and Hubble fluctuations. This is a new result that enhances the marginal discussion in \cite{CET} of the application of the CET proposal to LTB models. However, there are subtle but important differences between the various proposals: the condition for positive entropy production from the CET proposal is a local, necessary and sufficient, condition, while the conditions for the HBp and HBq proposals are domain dependent, and thus non-local, though they are expressible in terms of local fluctuations as sufficient (not necessary) conditions.  

While the conditions for entropy production are appealing and elegant, it is necessary to verify their actual fulfilment on the models. We undertake this task by means of analytic relations that hold at various asymptotic limits characteristic of the models: near the Big Bang, asymptotic time range, maximal expansion, near collapse and radial asymptotic regime. By collecting all this information we are able to provide a roughly consistent qualitative description of the full time behaviour of the entropy and entropy production valid for the three  proposals. This description reveals that entropy production for the three entropies is non--negative for all domains and regions of generic LTB models in which the density growing mode is dominant. As a consequence, entropy production is necessarily negative in the early time evolution of all models with a non-zero decaying mode (which is always dominant near a non--simulteneous Big Bang). These results can be connected with those of theoretical studies by Goode and Wainwright \cite{GooWain} and Lim {\it et al} \cite{Limetal} of the early time behavior and initial singularities in generic inhomogeneous models, and also with those obtained in a recent article by Bolejko and Stoeger \cite{bolstoeg}, who undertook a numerical study of entropy production (from various conceptual proposals) for spherically symmetric spacetimes with non-zero pressure and viscosity (we compare our results with those of these articles in section 10).

The content of the article is given as follows. In section 2, we review the definitions of the CET and the original HB entropy proposals. Section 3 is devoted to the presentation of LTB spacetimes in terms of the q--scalars, their fluctuations and perturbations. We provide a full derivation in section 4 of the conditions for positive entropy production for all proposals. In section 5 we comment on the subtle differences between these entropy production conditions: their `local' vs `non--local' and `necessary and sufficient' vs `only sufficient' nature. We examine in section 6 the fulfilment of these conditions in the asymptotic ranges of the time evolution of the models, looking at the general case (both density growing and decaying modes are non-zero), as well as the cases when one either one of these modes is suppressed. We use in section 7 the information obtained in section 6 to describe qualitatively the time evolution of the entropies, and present in section 8 three numerical examples that fully corroborate this qualitative description for the CET entropy. In section 9 we examine the integrability conditions of the CET entropy (which is defined through a Gibbs one--form), as well the radial asymptotic behaviour of the three entropies. We present a full discussion and summary of our results in section 10. The relation between the q--scalars and the traditional LTB variables is discussed in Appendix A, analytic solutions of the Friedman equation are summarized in Appendix B, Appendix C provides a brief discussion of the evolution equations used for the numerical examples. Formal results on the convergence of the HBp and HBq entropy functionals used in section 7 are proven in Appendix D. 

\section{The gravitational entropy.}

In this section, we will briefly present the two entropy proposals studied in this paper. We refer the reader to the original papers for more details.

\subsection{The  Clifton, Ellis and Tavakol (CET) entropy}

In the CET entropy proposal \cite{CET} a gravitational entropy $\sgr$ is constructed from the ``free'' gravitational field by demanding that it complies with basic consistency criteria, namely: that $\sgr$ and its associated entropy production are non--negative and that it is compatible with the Beckenstein--Hawking area formula when applied to black holes. For this purpose CET consider the Bell--Robinson tensor, $T_{abcd}$, which is the only totally symmetric traceless tensor that can be constructed with the conformal Weyl tensor $C_{abcd}$. 

However, since $T_{abcd}$ is fourth order (and thus its dimensions are $1/\hbox{cm}^4$), CET  consider a ``square root''  procedure expressing it as an irreducible algebraic decomposition in terms of a symmetric traceless second order tensor $t_{ab}$, which allows for a derivation of an ``effective'' or ``super'' energy--momentum tensor ${\cal T}_{ab}$ associated with the free gravitational field, with ``gravitational'' energy--momentum fluxes (energy density $\rhogr$, pressure $\pgr$, anisotropic stresses $\Pigr$ and heat flux $\qgr$) constructed by invariant contractions with the matter 4--velocity $u^a$ and projector $h_{ab}=u_au_b+g_{ab}$. A self consistent form for the ``gravitational entropy'' emerges by analogy with standard laws of fluid thermodynamics applied to the quantities associated with ${\cal T}_{ab}$. CET consider two paradigmatic types of gravitational fields, for which ${\cal T}_{ab}$ takes simple forms in terms of the Newman--Penrose conformal invariants $\Psi_2$ and $\Psi_4$: the ``Coulomb--like'' Petrov type D and the ``wave--like'' Petrov type N fields (see \cite{CET} for further details). 

For Coulomb--like fields CET obtain the following effective tensor and associated fluxes:
\ba \fl  \frac{{\cal T}^{ab}}{8\pi}=\alpha|\Psi_2| \left[x^ax^b+y^ay^b-2\left(z^az^b-u^au^b\right)\right]=\rhogr u^au^b + \pgr h^{ab}+2q^{(a}_{\tiny{\textrm{grav}}} u^{b)}+\Pigr,\nonumber\\
\fl 8\pi\rhogr=2\alpha|\Psi_2|,\quad \pgr=\qgr=0,\quad 8\pi\Pigr=\frac{\alpha|\Psi_2|}{2}(x^ax^b+y^ay^b-z^az^b+u^au^b),\nonumber\label{effective}\\
\ea
where $\alpha$ is a positive constant to provide the appropriate physical units and $\{x^a,\,y^a,\,z^a,\,u^a\}$ is an orthonormal tetrad. By analogy with the off--equilibrium Gibbs equation (see \cite{rund}) CET obtain the following expression for the entropy production:
\begin{equation}\Tgr\dot\sgr = (\rhogr V)\dot{}=-V\sigma_{ab}\left[\Pigr+\frac{4\pi(\rho+p)}{3\alpha|\Psi_2|}E^{ab}\right],\label{gibbs}\end{equation}
where $V$ is a suitable local volume, $E^{ab}=u_au_b C^{acbd}$ is the electric Weyl tensor and the ``gravitational'' temperature $\Tgr$ is given by
\begin{equation} \Tgr = \frac{\left|\dot u_az^a+\theta/3+\sigma_{ab}z^az^b\right|}{2\pi},\label{Tgr}  \end{equation}
where $\dot u_a =u^b\nabla_a u_b$ is the 4--acceleration and $\theta\equiv \tilde\nabla_cu^c=h_c^b\nabla_b u^c$ is the isotropic expansion scalar. As commented by CET, the terms inside the brackets in the right hand side of (\ref{gibbs}) play the role of ``effective'' relativistic dissipation terms in analogy with dissipative matter sources, though this is merely an analogy, and since the actual sources are conserved, the Gibbs equation (\ref{gibbs}) does not imply that they exchange energy or momentum with the free gravitational fields associated with (\ref{effective}). On the other hand, CET justify $\Tgr$ in (\ref{Tgr}) as a local temperature that reduces to the semi--classical Unruh and Hawking temperatures in the appropriate limits (see \cite{CET}).       

\subsection{The Hosoya--Buchert (HB) entropy}

The HB entropy is originally inspired by the relative information entropy, also known as the Kullback-Leibler divergence \cite{Info}, defined as:
$$
s_{\textrm{\tiny{KL}}}=\sum_{i}p_{i}\ln\left[\frac{p_{i}}{{\cal P}_{i}}\right],\label{KL}
$$
where $p_{i}$ is the actual probability density of the random variable $i$, and ${\cal P}_{i}$ the expected one. If the two probability densities coincide for all the variables (i.e. if our information on the system is complete), then $s_{\textrm{\tiny{KL}}}=0$. On the other hand, if, for at least one of the variables, $p_{i}\neq {\cal P}_{i}$, then in general, $s_{\textrm{\tiny{KL}}}\neq 0$. Hosoya and Buchert constructed in \cite{HB1,HB2,HB3} an entropy based on the Kullback-Leibler divergence, suitable to study the emergence of inhomogeneities in cosmological models. They defined what we shall denote by the Hosoya-Buchert (HB) entropy:
\begin{equation}
\shb=\int_{{\cal D}}\rho\ln\left[\frac{\rho}{\langle\rho\rangle_{\cal D}}\right]\dd\mu,\label{sHB}
\end{equation}
where $\rho$ is the relativistic energy density and $d\mu$ is the Riemannian measure on the spacelike hypersurfaces on which the Buchert's average $\langle\,.\,\rangle_{\cal D}$ of $\rho$:
\begin{equation}\langle\rho\rangle_{{\cal D}}=\frac{\int_{{\cal D}}{\rho\,\dd\mu}}{\int_{{\cal D}}{\dd\mu}},\label{rhoav}\end{equation}
is computed over an averaging domain ${\cal D}$. Evidently, an entropy measure like that of HB can also be defined with any self--consistent energy density average . The entropy measure (\ref{sHB}) can also be defined with a scalar weighed average (q--average) (see \cite{part1}). We provide rigorous proof in Appendix D.1 that $\shb$ is non--negative for every scalar averaging formalism and for every domain as long as $\rho\geq 0$.  

\section{LTB dust models in the q--scalar representation.}

We shall describe LTB dust models in the following useful FLRW--like metric parametrization
\footnote{The relation between this metric parametrization and the standard metric form and variables of the models is discussed in detail in Appendix A. See \cite{part1,part2} for a comprehensive discussion on the q--scalar representation introduced in this section.} 
\ba \fl \dd s^2 =-\dd t^2+ a^2\left[\frac{\Gamma^2}{1-\KK_{q0}r^2}\dd r^2+r^2\left(\dd\vartheta^2+\sin^2\theta\dd\varphi^2\right)\right],\label{ltb2}\\
\fl a=a(t,r),\qquad \dot a^2 =\left[\frac{\partial a}{\partial t}\right]^2=\frac{8\pi\rho_{q0}}{3a}-\KK_{q0},\qquad \Gamma=1+\frac{ra'}{a},\qquad a' =\frac{\partial a}{\partial r},\label{aGdef}\ea 
where $\KK_{q0}=\KK_q(t_0,r)$ and $\rho_{q0}=\rho_q(t_0,r)$ are defined further ahead (see equation (\ref{rhoKK})) and the subindex ${}_0$ will denote henceforth evaluation at an arbitrary fiducial hypersurface $t=t_0$. We remark that $a_0=\Gamma_0=1$ (see Appendix A). 

It is useful to describe the dynamics of the models by means of their covariant objects given in terms of the representation of ``q--scalars'' and their perturbations (see \cite{part1,part2,sussmodes} for a comprehensive discussion on the definition and properties of these perturbations). For every LTB scalar $A$, the associated q--scalar $A_q$, perturbation $\Da_q$ and fluctuation $\Del_q(A)$ are defined by the correspondence rules  
\ba A_q =\frac{\int_0^r{A\,a^3\,\Gamma\,\bar r^2\,\dd\bar r}}{\int_0^r{a^3\,\Gamma\,\bar r^2\,\dd\bar r}}=\frac{3\int_0^r{A\,a^3\,\Gamma\,\bar r^2\,\dd\bar r}}{a^3r^3},\label{Aqdef}\\
\Da_q =\frac{A-A_q}{A_q} = \frac{rA'_q/A_q}{3\Gamma}=\frac{1}{r^3 a^3 A_q}\int_0^r{A'\,\bar r^3\, a^3\dd \bar r},\label{Dadef}\\
\Del_q(A)\equiv A-A_q=A_q\Da_q=\frac{r A'_q}{3\Gamma}=\frac{1}{r^3\,a^3}\int_0^r{A'\, \bar r^3\,a^3\dd \bar r}.\label{Deldef}
\ea
where (\ref{Dadef}) and (\ref{Deldef}) follow directly by differentiation and integration by parts of (\ref{Aqdef}) and allow, in general, to compute $\Da_q$ and $\Del_q(A)$ in terms of the gradients $A'_q$ and the scale factor $\Gamma$. The q--scalars are directly related to proper volume averages with weight factor $\FF=\sqrt{1-\KK_{q0}r^2}$ (see \cite{part1,part2} for further details and explanations).   

The basic LTB covariant scalars are: (i) the rest--mass density $\rho$, (ii) the Hubble scalar $\HH\equiv \theta/3$ and (iii) the spatial curvature scalar $\KK\equiv \RR/6$ (with $\RR$ the Ricci scalar of hypersurfaces $t=$ const.). In the q--scalar representation these scalars take the forms of exact perturbations \cite{part1,part2}:
\begin{equation}\fl  \rho=\rho_q(1+\Drho_q),\qquad \HH=\HH_q(1+\Dh_q),\qquad \KK=\KK_q(1+\DKK_q),\label{rhoHHKK}\end{equation}
with their associated q--scalars and perturbations given by \cite{sussmodes}: 
\ba \fl \frac{8\pi}{3}\rho_q =\frac{8\pi}{3}\frac{\rho_{q0}}{a^3}=\frac{\Omega_{q0}\HH_{q0}^2}{a^3},\qquad \KK_q=\frac{\KK_{q0}}{a^2}=\frac{(\Omega_{q0}-1)\HH_{q0}^2}{a^2},\label{rhoKK}\\
\fl \HH_q =\frac{\dot a}{a}=\frac{\left[\frac{8\pi}{3}\rho_{q0}-\KK_{q0}a\right]^{1/2}}{a^{3/2}}=\frac{\HH_{q0}\left[\Omega_{q0}-(\Omega_{q0}-1)a\right]^{1/2}}{a^{3/2}},\label{HHq}\\
\fl \Drho _q= \frac{\Jg+\Jd}{1-\Jg-\Jd}=\frac{1+\Drho_0}{\Gamma}-1,\label{Drho}\\
\fl  \DKK_q = \frac{2\,(\Jg+\Jd-\Dig)}{3(1-\Jg-\Jd)}=\frac{2/3+\DKK_0}{\Gamma}-\frac{2}{3},\label{DKK}\\
\fl \Dh_q= \frac{(2+\Omega_q)(\Jg+\Jd)-2(1-\Omega_q)\Dig}{6(1-\Jg-\Jd)}=\frac{\Omega_q}{2}\Drho-\frac{\Omega_q-1}{2}\DKK,\label{Dh}\ea
where the q--scalar $\Omega_q$ and its perturbation are given by  
\begin{equation}\fl \Omega_q \equiv \frac{8\pi\rho_q}{3\HH_q^2}=\frac{\Omega_{q0}}{\Omega_{q0}+(1-\Omega_{q0})a},\qquad \DOm_q = \frac{(1-\Omega_q)\,\left(\Jg+\Jd+2\Dig\right)}{3(1-\Jg-\Jd)},\label{OmDOm}\end{equation}
and $\Jg,\,\Jd$ are the exact generalizations of the growing and decaying density modes of linear perturbation theory (see \cite{sussmodes}):
\ba \Jg = 3\Dig\left[\HH_q (t-\tbb)-\frac{2}{3}\right],\qquad \hbox{density growing mode},\label{gmode}\\
\Jd = 3\Did\,\HH_q,\qquad\qquad\qquad \hbox{density decaying mode}, \label{dmode}\ea
where $\tbb=\tbb( r)$ is the Big Bang time,  $\Dig$ and $\Did$ (both assumed non-zero unless stated otherwise) are the ``amplitudes'' of the modes:
\ba \fl \Dig \equiv \frac{\Drho_{q0}-\frac{3}{2}\DKK_{q0}}{1+\Drho_{q0}},\label{Dig}\\ \fl \Did \equiv -\frac{\HH_{q0}(t_0-\tbb)\left[\Drho_{q0}-\frac{3}{2}\DKK_{q0}\right]+\DKK_{q0}-\Drho_{q0}}{\HH_{q0}(1+\Drho_{q0})}=\frac{r\,\tbb'}{3(1+\Drho_{q0})} \label{Did}\ea
which can also be given in terms of $\Dh_{q0},\,\DOm_{q0}$ by the relations
\begin{equation} \Drho_{q0}=\DOm_{q0}+2\Dh_{q0},\qquad  \DKK_{q0}=2\Dh_{q0}-\frac{\Omega_{q0}\DOm_{q0}}{1-\Omega_{q0}}.\label{ivpconstr} \end{equation}
Given the constraints among the basic initial q--scalars, $A_{q0}=\rho_{a0},\,\HH_{q0},\,\KK_{q0},\,\Omega_{q0}$, and the relation between their gradients and the perturbations $\Da_{q0}=(r/3)A'_{q0}/A_{q0}$ that follows from (\ref{Dadef}), any LTB model can be uniquely specified by selecting as free parameters (initial conditions) any two of the four $A_{q0}$. The dynamics of the models becomes fully determined, either analytically through the scaling laws (\ref{rhoHHKK})--(\ref{ivpconstr}) once we have add the analytic expression relating $\HH_q(t-\tbb)$ with $a$ (see (\ref{HTq})--(\ref{H32}) in Appendix B), or numerically by solving the appropriate systems of evolution equations (see Appendix C and examples in \cite{part2,sussmodes}). We shall use for various calculations and graphs in the remaining of this paper both the numerical and analytic approaches. 

The main LTB proper tensors are the shear tensor ($\sigma_{ab}=\tilde\nabla_{(a}u_{b)}-\HH h_{ab}$), the electric Weyl tensor ($E_{ab}=u_au_b C^{acbd}$) and Weyl tensor ($C^{acbd}$). These tensors and their eigenvalues take the form:
\bse\ba \fl \sigma_{ab}=\Sigma\,\hbox{\bf{e}}_{ab},\qquad E_{ab}=\Psi_2\,\hbox{\bf{e}}_{ab},\qquad C^{ab}_{cd}=\Psi_2\,\left(h^{[a}_{[c}-3u_{[c}u^{[a}\right)\,\hbox{\bf{e}}^{b]}_{d]},\label{sigEC}\\
\fl \Sigma =\frac{1}{6}\hbox{\bf{e}}_{ab}\sigma^{ab}=-\frac{\dot\Gamma}{3\Gamma}=-\Del_q(\HH),\qquad \Psi_2 =\frac{1}{6}\hbox{\bf{e}}_{ab}E^{ab}=\frac{4\pi}{3}\,\Del_q(\rho),\label{SigPsi2}\ea\ese
where $\Psi_2$ is the Petrov type D conformal invariant in (\ref{effective}) and $\hbox{\bf{e}}^a_b=h^a_b-3n^a n_b$,  with $n_a=\sqrt{g_{rr}}\delta^r_a$, is the common covariantly constant tensor of Petrov type D LRS spacetimes \cite{LRS}. 
            
\section{Gravitational entropy in LTB dust models.}

\subsection{The CET gravitational entropy.}

From the expressions associated with the effective energy--momentum tensor (\ref{effective}) and the entropy production law (\ref{gibbs}) and gravitational temperature (\ref{Tgr}), CET obtain the following entropy production law for LTB models:
\begin{equation} \Tgr\dot\sgr=\partial_{t}(\rhogr V),\label{CET2} \end{equation}
where $V=\ell^3 = a^3\Gamma$ is the local volume defined by the condition $\HH =\dot\ell/\ell=(a^3\Gamma)\dot{}/(3a^3\Gamma)$ and $\rhogr,\,\Tgr$ are given by
\ba
\fl 8\pi\rhogr &=&2\alpha\frac{|M-4\pi R^{3}\rho/3|}{R^{3}}=2\alpha|\Psi_2|=\frac{8\pi\alpha}{3}|\Del_q(\rho)|=\frac{8\pi\alpha}{3}\rho_q|\Drho_q|,\label{rhograv2}\\
\fl\Tgr&=&\frac{1}{2\pi}\left|\frac{\dot{R}^{'}}{R^{'}}\right|=\frac{|\HH_q|\,|\, 1+3\Dh_q\,|}{2\pi}.\label{Tgrav2}
\ea
where $R=a r$, and we used (\ref{ivf1}), (\ref{basic}), (\ref{qbasic}) and (\ref{SigPsi22}) to express $\rho,\,M/R^3$ and $\dot R'/R'$ in terms of q--scalars and their fluctuations. Inserting (\ref{rhograv2}) and (\ref{Tgrav2}) into (\ref{CET2}) we obtain:
\begin{equation}
\fl \dot\sgr=\frac{2\pi\alpha}{3}\frac{\partial_{t}\left(\rho_{q}a^3\Gamma\left|\Drho\right|\right)}{|\HH_{q}||1+3\Dh_q|}= \frac{2\pi\alpha\rho_{q0}}{3}\frac{\partial_{t}\left(\Gamma\left|\Drho\right|\right)}{|\HH_{q}||1+3\Dh_q|}\label{CETc12}.
\end{equation}
which determines the sign of $\dot\sgr$:
\begin{equation}
\dot\sgr\geq 0 \quad \Leftrightarrow \quad \partial_{t}\left[\Gamma|\Drho_q|\right]\geq 0.\label{CETc22}
\end{equation}
From (\ref{rhoHHKK}) and (\ref{SigPsi2}) we obtain after some calculations:
\ba \fl \partial _t\left(\Gamma\,|\Drho_q|\right) = \dot\Gamma|\Drho|+\Gamma \partial_t(|\Drho|)=\left[\partial_t(\,|\Drho_q|\,)+3\HH_q|\Drho_q|\Dh_q\right]\Gamma,\ea
where  we used $\dot \Gamma =3\Gamma \Del_q(\HH)=3\Gamma\HH_q\Dh$ (from (\ref{SigPsi2})) and we assume henceforth that $\Gamma>0$ holds to avoid shell crossing singularities \cite{RadAs,RadProfs,sussmodes}.  Considering that
\ba\fl |\Drho_q| = \left\{\begin{array}{c}
\Drho_q\quad\hbox{if}\quad \Drho_q> 0,\\
0 \quad\hbox{if}\quad \Drho_q= 0,\\
-\Drho_q \quad\hbox{if}\quad \Drho_q< 0,  
\end{array}\right.\quad \partial_t(\,|\Drho_q|\,) = \left\{\begin{array}{c}
\partial_t(\,\Drho_q\,)=\dot\Drho_q\quad\hbox{if}\quad \Drho_q> 0,\\
0 \qquad\qquad\quad\hbox{if}\quad \Drho_q= 0,\\
\partial_t(-\Drho_q)=-\dot\Drho_q \quad\hbox{if}\quad \Drho_q< 0,  
\end{array}\right.\nonumber\\
\ea
and using the evolution equation (19a) of \cite{part2} to eliminate $\dot\Drho_q$, we obtain 
\begin{equation}\fl \partial _t\left(\Gamma\,|\Drho|\right) = \left\{\begin{array}{c}
-3\HH_q\Dh_q=-3\HH_q\Dh_q\Drho_q/|\Drho_q| \quad\hbox{if}\quad \Drho_q> 0,\\
0 \qquad\qquad\qquad\qquad\qquad\qquad \hbox{if}\quad \Drho_q= 0,\\
3\HH_q\Dh_q=3\HH_q\Dh_q\Drho_q/|\Drho_q| \quad\hbox{if}\quad \Drho_q< 0.  
\end{array}\right.\label{CETc122}\end{equation}
where we used the fact that $\Drho_q/|\Drho_q|=\hbox{signum}(\Drho)$. From (\ref{Deldef}), and since $\rho_q\geq 0$ implies that $\Drho_q$ and $\Del_q(\rho)$ have the same sign, we can now express condition (\ref{CETc22}) in terms of the fluctuations $\Del_q(\HH)$ and $\Del_q(\rho)$
\begin{equation}
\dot\sgr\geq 0 \quad \Leftrightarrow \quad \left\{\begin{array}{c}
\Del_q(\HH)<0\quad\hbox{if}\quad \Del_q(\rho)> 0,\\
\Del_q(\HH)=0 \quad\hbox{if}\quad \Del_q(\rho)= 0,\\
\Del_q(\HH)>0 \quad\hbox{if}\quad \Del_q(\rho)< 0,  
\end{array}\right..\label{CETc23}
\end{equation}
or equivalently from (\ref{Deldef}): 
\begin{equation} \fl \dot\sgr\geq 0 \quad \Leftrightarrow \quad \Del_q(\rho)\Del_q(\HH)=\frac{\rho'_q\HH'_q}{(3\Gamma/r)^2}=\frac{\int_0^r{\rho'\,(a\bar r)^3\dd\bar r}\int_0^r{\HH'\,(a\bar r)^3\dd\bar r}}{R^6}\leq 0,\label{CETcond}\end{equation}
which is reminiscent to a condition of negative correlation between fluctuations of the energy density and Hubble scalar.

We remark, from (\ref{SigPsi2}), that the non--negative entropy production condition (\ref{CETcond}) can also be written as:
\begin{equation} \fl \dot\sgr\geq 0 \quad \Leftrightarrow \quad \Sigma\,\Psi_2 \geq 0\qquad \hbox{or}\qquad  \dot\sgr\geq 0 \quad \Leftrightarrow \quad \sigma_{ab}E^{ab}\geq 0\label{CETcond2}\end{equation}
where $\Sigma$ and the conformal invariant $\Psi_2$ are, respectively, the eigenvalues of the shear and electric Weyl tensors. Another alternative form for (\ref{CETcond}) is 
\begin{equation} \dot\sgr\geq 0 \quad \Leftrightarrow \quad \Del_q(\RRR)\,\Del_q(\HH) \leq 0,\end{equation} 
where $\RRR$ is the (4--dimensional) Ricci scalar and we used the fact that $\RRR=8\pi\rho$, and thus $\Del(\RRR)=8\pi\Del(\rho)$, hold for LTB models.  

\subsection{The HBp and HBq gravitational entropies.}

For a fixed arbitrary spherical domain $\DD[r_b]$ centered on $r=0$, whose proper volume is 
\begin{equation} \fl \VV_p[r_b]= \int_{\DD[r]}{\dd\VV_p}=4\pi\int_0^{r_b}{\FF^{-1}a^3\Gamma \bar r^2\dd\bar r},\qquad \FF\equiv \sqrt{1-\KK_{q0}r^2},\label{propvol}\end{equation}
the original HB entropy functional (\ref{sHB}) for Buchert's average (\ref{rhoav}) (to be denoted henceforth as the ``HBp'' entropy) applied to LTB models takes the form:
\footnote{We will use the subindex ${}_p$ to emphasize the connection to Buchert's proper volume average, using the notation ``$\Aav_p[r_b]$'' as domain indicator instead of the usual form $\Aav_\DD$. The subindex ${}_p$ will be attached to all quantities related to this average, such as the local and non--local fluctuations $\Del_p(A)$ and $\Denl_p(A)$ \cite{part2}, and their associated local functions, $A_p$, to distinguish them from the analogous objects constructed with the correspondence rule (\ref{Aqdef}): the q--scalars, the q--average and their local and non--local fluctuations and perturbations, all of which carry the subindex ${}_q$.}
\ba \fl \shb{}_p[r_b] -\shb{}_p^{(\textrm{\tiny{eq}})}= \gamma_0\int_{\DD[r_b]}{\rho\ln\left[\frac{\rho}{\rhoav_p[r_b]}\right]\dd\VV_p}=\gamma_0\VV_p[r_b]\left\langle \rho\ln\left[\frac{\rho}{\rhoav_p[r_b]}\right]\right\rangle_p[r_b],\nonumber\\\label{sHB2}\ea
where $\shb{}_p^{(\textrm{\tiny{eq}})}>0$ is the ``equilibrium'' entropy, $\gamma_0$ is a constant so that the left hand side of (\ref{sHB2}) has units of entropy, while the density average is
\begin{equation}  \rhoav_p[r_b]=\frac{\int_{\DD[r]}{\rho\,\dd\VV_p}}{\int_{\DD[r]}{\dd\VV_p}} =\frac{M_p[r_b]}{\VV_p[r_b]}\label{rhoave2}\end{equation}
where $M_p$ is the domain's proper mass--energy function
\begin{equation} M_p[r_b]=\int_{\DD[r]}{\rho\,\dd\VV_p}=\int_0^{r_b}{\rho \VV'_p \dd r}=4\pi\int_0^{r_b}{\rho\, \FF^{-1}a^3\Gamma \bar r^2\dd\bar r},\label{Mp}\end{equation}
which is independent of $t$ (since $\rho a^3\Gamma=\rho_0$ follows from (\ref{rhoHHKK}), (\ref{rhoKK}) and (\ref{Drho})). Following \cite{HB1,HB2,HB3} we evaluate $\dot \shb{}_p$ by applying to (\ref{sHB2}) the time derivative commutation rule for any averaged scalar (we omit the domain indicator $[r_b]$ to simplify notation):
\ba \fl  \Aav\dot{}_p-\langle\dot A\rangle_p =3\left[\langle A\HH\rangle_p-\Aav_p\HHav_p\right]=3\langle (A-\Aav_p)(\HH-\HHav_p)\rangle_p=\hbox{{\bf Cov}}_p(A,\HH),\nonumber\\
\label{tder}\ea
which yields after some algebraic manipulation:
\ba \fl \frac{\dot \shb{}_p}{\gamma_0\VV_p}&=&-\langle \dot\rho\rangle_p+\rhoav\dot{}_p=-3\left[\langle \rho\HH\rangle_p-\rhoav_p\HHav_p\right]=- 3\langle \Denl_p(\rho)\Denl_p(\HH)\rangle_p\nonumber\\
\fl &=&-\hbox{{\bf Cov}}_p(\rho,\HH),\label{Sdot1}\ea
where 
\begin{equation}\Denl_p(\rho)=\rho( r)-\rhoav_p[r_b],\qquad \Denl_p(\HH)=\HH( r)-\HHav_p[r_b]\end{equation}
are the non--local fluctuations 
\footnote{The $\Denl_p(A)$ and their analogues $\Denl_q(A)$ are non--local fluctuations because they depend on both inner points $r<r_b$ and on the boundary $r=r_b$ of $\DD[r_b]$ (notice that $0\leq r\leq r_b$, see \cite{part1,part2}). In contrast, the fluctuations $\Del_q(A)=A-A_q$ defined by (\ref{Deldef}) and (\ref{denlq}) are local because both $A$ and $A_q$ are evaluated for the same value of $r$ (the same holds for the analogous local fluctuations $\Del_p(A)=A-A_p$).}
of $\rho$ and $\HH$, while $\hbox{{\bf Cov}}_p$ denotes the covariance statistical moment (correlation) with respect to the involved Buchert's averages (we have removed the domain indicator $[r_b]$ to simplify notation). The necessary and sufficient condition for a positive HBp entropy production is then  
\begin{equation}\dot \shb{}_p[r_b]\geq 0 \quad \Leftrightarrow\quad \hbox{{\bf Cov}}_p[r_b](\rho,\HH)\leq 0,\label{HBcond}\end{equation}
which directly relates this entropy production to the negative statistical correlation of $\rho$ and $\HH$ in an arbitrary domain $\DD[r_b]$.

Instead of using Buchert's density average (\ref{rhoave2}) to define the gravitational entropy (\ref{sHB}) for LTB models, we may consider the quasi--local weighted average (q--average) for a domain $\DD[r_b]$ defined by the correspondence rule (\ref{Aqdef})
\footnote{Notice that $A_q( r)$ and $\Aav_q[r_b]$ are different objects even if both follow from the same correspondence rule (\ref{Aqdef}). The q--average $\Aav_q[r_b]$ is a functional and $A_q( r)$ is the function constructed from this functional by considering a varying domain boundary. Hence, $\Aav_q[r_b]$ is effectively a constant for inner points $r<r_b$ of $\DD[r_b]$, while $A_q( r)$ is locally varying for these points and both coincide at the boundary $r=r_b$ for every $r_b$ (see figure 1 of \cite{sussBR} and equation (\ref{AqAavq})). Analogous local functions $A_p$ are also defined for Buchert's average functional $\Aav[r_b]$ in (\ref{Apdef}) (see \cite{part1} and \cite{sussBR} for a comprehensive discussion).}
\begin{equation} \rhoav_q[r_b]=\frac{\int_{\DD[r_b]}{\rho\,\FF\,\dd\VV_p}}{\int_{\DD[r_b]}{\FF\,\dd\VV_p}}=\frac{M_q[r_b]}{\VV_q[r_b]},\label{rhoaveq2}\end{equation}
where $\dd \VV_q=\FF\dd\VV_p$ defines the quasi--local volume of the domain 
\begin{equation} \VV_q[r_b]=\int_{\DD[r_b]}{\dd\VV_q}=4\pi\int_0^{r_b}{a^3\,\Gamma\,\bar r^2\,\dd\bar r}=\frac{4\pi}{3}a^3(r_b)r_b^3.\label{qvol}\end{equation}
and $M_q$ is the quasi--local mass--energy function of the domain 
\begin{equation} M_q[r_b]=\int_\DD{\rho\,\dd\VV_q}=4\pi\int_0^{r_b}{\rho\,\VV'_q\dd \bar r}=4\pi\int_0^{r_b}{\rho\,a^3\,\Gamma\,\bar r^2\dd\bar r},\label{Mq}\end{equation}
which is (like $M_p$) independent of $t$ by virtue of (\ref{rhoHHKK}), (\ref{rhoKK}) and (\ref{Drho}). Following \cite{part1}, we define the ``HBq'' gravitational entropy for the domain $\DD[r_b]$ along the lines of the HBp entropy defined before:
\ba \fl \shb{}_q[r_b] -\shb{}_q^{(\textrm{\tiny{eq}})}= \gamma_0\int_{\DD[r_b]}{\rho\ln\left[\frac{\rho}{\rhoav_q[r_b]}\right]\FF\dd\VV_p}=\gamma_0\VV_q[r_b]\left\langle \rho\ln\left[\frac{\rho}{\rhoav_q[r_b]}\right]\right\rangle_q[r_b],\nonumber\\\label{sHBq}\ea
As shown in \cite{part1}, the definition (\ref{sHBq}) yields results that are equivalent to those obtained from (\ref{sHB2}):
\ba \fl \frac{\dot \shb{}_q[r_b]}{\gamma_0\VV_q[r_b]}=-3\left[\langle \rho\HH\rangle_q-\rhoav_q\HHav_q\right]=- 3\langle \Denl_q(\rho)\Denl_q(\HH)\rangle_q=-\hbox{{\bf Cov}}_q(\rho,\HH),\nonumber\\
\label{Sdot1q}\ea
so that
\begin{equation}\fl \dot \shb{}_q[r_b]\geq 0 \quad \Leftrightarrow\quad \hbox{{\bf Cov}}_q[r_b](\rho,\HH)= \langle\Denl_q(\HH)\Denl_q(\rho)\rangle_q[r_b]\leq 0,\label{HBcondq}\end{equation}
where
\begin{equation}\fl \Denl_q(\rho)=\rho( r)-\rhoav_q[r_b],\qquad \Denl_q(\HH)=\HH( r)-\HHav_q[r_b]\label{denlq}\end{equation}
are the non--local fluctuations with respect to the q--averages $\rhoav_q[r_b]$ and $\HHav_q[r_b]$. 

\section{Local vs non--local entropies.} 

It is quite interesting that the condition for a positive entropy production from the CET proposal (\ref{CETcond}) resembles that obtained from the HBp proposal (\ref{HBcond}) and its quasi--local version HBq  in (\ref{HBcondq}). In fact, the resemblance is more striking if we consider the HBq entropy because q--scalars and q--averages coincide at the domain boundary (see \cite{part1} and figure 1 of \cite{sussBR} for a comprehensive discussion):
\begin{equation}A_q( r_b) = \Aav_q[r_b]\quad \Rightarrow\quad \Denl_q(A)|_{r=r_b}=\Del_q(A)|_{r=r_b}.\label{AqAavq}\end{equation}
Nevertheless, whether we consider Buchert's average or the q--average, it is important to remark that there are important but subtle differences between (\ref{CETcond}) and either one of (\ref{HBcond}) or (\ref{HBcondq}): the CET entropy is defined for a local volume $V$, which is consistent with the fact that the $\Del_q$ fluctuations are local, and thus (\ref{CETcond}) is a local condition evaluated in a point--wise manner, whereas the $\Denl_p$ and $\Denl_q$ fluctuations are non--local, and thus the HB and HBq entropies in (\ref{HBcond}) and (\ref{HBcondq}) must be evaluated though proper volume averaging over a domain $\DD[r_b]$. 

This difference is important, since (\ref{CETcond}) is necessary and sufficient by definition, as it follows directly from the original CET article \cite{CET}) and its fulfillment can be tested by local evaluation of the involved quantities (we only need to evaluate both fluctuations $\Del_q(\rho)$ and $\Del_q(\HH)$ at each point). Conditions (\ref{HBcond}) and (\ref{HBcondq}) are also necessary and sufficient by definition (see references \cite{HB1,HB2,HB3,part1}), but their necessary and sufficient nature is strictly domain dependent, that is: it only applies after the integrals in the involved averages have been evaluated for any given domain. As a consequence, we cannot rule out that either one of the latter conditions (say (\ref{HBcond})) may hold ({\it i.e.} $\hbox{{\bf Cov}}_p(\rho,\HH)\leq 0$) even if $\Del_p(\rho)\Del_p(\HH)\leq 0$ fails to hold in inner points $r<r_b$ of $\DD[r_b]$ (the same situation occurs for $\hbox{{\bf Cov}}_q(\rho,\HH)$). 

However, we can obtain weaker conditions that are only sufficient (and not necessary) by looking for sign conditions in the integrands {\it before} the evaluation of the integrals: if these conditions are fulfilled the integrals (once evaluated) will have the desired sign, but the converse statement is false. These weaker conditions follow from the fact that if $\Denl_p(A)\Denl_p(B)\leq 0$ holds for every $0\leq r\leq r_b$, it implies $\hbox{{\bf Cov}}_p(A,B)\leq 0$ for every pair of scalars $A,\,B$ (the same occurs with the $\Denl_q$), which leads to the following sufficient but not necessary conditions:
\bse\ba \Denl_p(\rho)\Denl_p(\HH)\leq 0 \quad \forall\;\; 0\leq r\leq r_b\quad \Rightarrow\quad \dot\shb{}_p(r_b) \geq 0,\\
\Denl_q(\rho)\Denl_q(\HH)\leq 0  \quad\forall\;\; 0\leq r\leq r_b\quad \Rightarrow\quad \dot\shb{}_q(r_b) \geq 0, \label{NLconds}\ea\ese 
which must still be evaluated at every point in $\DD[r_b]$. More useful sufficient (not necessary) conditions on $\dot\shb{}_p$ and $\dot \shb{}_q$ can be obtained that only involve evaluating local fluctuations at the boundary of each domain. For this purpose, we define the local fluctuations equivalent to the quasi--local fluctuations (\ref{Deldef}):
\begin{equation}\Del_p(A)=A-A_p = \frac{A'_p}{\VV'_p/\VV_p}=\frac{1}{\VV_p}\int_0^r{A'\,\VV_p\,\dd\bar r},\label{Apdef}\end{equation}
where $A_p$ are the local functions whose correspondence rule is the same as that of Buchert's average ({\it i.e.} equation (\ref{rhoave2})). The following results on quadratic fluctuations proven in \cite{part1} and \cite{sussBR}
\bse\ba \langle \Denl_p(\rho)\Denl_p(\HH)\rangle_p[r_b]=\langle\Del_p(\rho)\Del_p(\HH)\rangle_p[r_b],\label{result1}\\
\langle \Denl_q(\rho)\Denl_q(\HH)\rangle_q[r_b]=\langle\Del_q(\rho)\Del_q(\HH)\rangle_q[r_b],\label{result2}\ea\ese
yield the following sufficient (not necessary) conditions
\bse\ba \left[\,\Del_p(\rho)\Del_p(\HH)\,\right]|_{r=r_b}\leq 0 \quad \Rightarrow\quad \dot\shb{}_p(r_b) \geq 0,\label{HBcond01}\\
\left[\,\Del_q(\rho)\Del_q(\HH)\,\right]|_{r=r_b}\leq 0 \quad \Rightarrow\quad \dot\shb{}_q(r_b) \geq 0, \label{HBcond01}\ea\ese 
which are valid for every domain, and since $r=r_b$ is arbitrary, they can be evaluated locally for all the range of $r$ and can be stated simply as local sufficient conditions: 
\bse\ba \Del_p(\rho)\Del_p(\HH)\leq 0 \quad \Rightarrow\quad \dot\shb{}_p \geq 0,\label{HBcond1}\\
\Del_q(\rho)\Del_q(\HH)\leq 0 \quad \Rightarrow\quad \dot\shb{}_q \geq 0. \label{HBcond2}\ea\ese
This is an important result, as it shows that the conditions for entropy production that emerge from the CET proposal and the HBq proposal are essentially the same, as both are based on q--scalars: compare (\ref{CETcond}) and (\ref{HBcond2}), the only difference being that they are necessary and sufficient for the CET proposal and sufficient but not necessary for the HBq proposal. 

The conditions from the HBp proposal (with Buchert's average) are also the same as those from CET, but the fluctuations in the HBp case must be evaluated from the local functions $A_p$ in (\ref{Apdef}) associated with Buchert's average and that are analogous to $A_q$. However, we can readily obtain sufficient (not necessary) conditions for (\ref{CETcond}) and (\ref{HBcond2}) that are valid for both averages under certain restrictions: if both $\rho$ and $\HH$ are monotonous in a given domain $\DD[r_b]$, then
\begin{equation}\fl \hbox{if}\quad\rho'\HH'\leq 0\quad\hbox{holds}\quad \Rightarrow\quad \dot\shb{}_p[r_b] \geq 0\quad\hbox{{\bf and}}\quad\dot\shb{}_q[r_b] \geq 0\quad \hbox{hold},\label{HBcond3}\end{equation}
where we used the fact that the integrals in (\ref{Deldef}) and (\ref{Apdef}) for $A=\rho,\,\HH$ involve $\rho'$ and $\HH'$ as integrands. Hence, if these gradients have opposite signs in all points of a domain $\DD[r_b]$ the products of local fluctuations $\Del_q(\rho)\Del_q(\HH)$ and $\Del_p(\rho)\Del_p(\HH)$ will be negative (but the converse is false).  

\section{Probing the conditions for entropy production in the asymptotic limits.}

While the conditions for positive entropy production from the CET and HB proposals are elegant and plausible, it is necessary to verify their actual fulfilment in generic regular LTB models, at least in the asymptotic ranges of their time evolution. We assume henceforth that shell crossings are absent ($\Gamma>0$ holds for $a>0$, see the conditions for this in \cite{sussmodes}). We examine first the case of the CET entropy in (\ref{CETcond}), which is a sufficient but not necessary condition for the HBq case in (\ref{HBcondq}) and (under certain restrictions) for the HB case in (\ref{HBcond}). 

In order to examine qualitatively the fulfilment of (\ref{CETcond}) and (\ref{HBcondq}) in various evolution ranges we remark that
\begin{equation} \Del_q(\rho)\Del_q(\HH)=\rho_q\,\HH_q\,\Drho_q\,\Dh_q.\label{deltacond}\end{equation}
where, following \cite{sussmodes}, we will use the expressions for the perturbations $\Drho_q,\,\Dh_q$ given by (\ref{Drho}) and (\ref{Dh}) in terms of the exact generalizations of the growing/decaying density modes (\ref{gmode})--(\ref{ivpconstr}) and the exact analytic forms for $t-\tbb,\,\HH_q$ and $\HH_q(t-\tbb)$ given by (\ref{hypsol})--(\ref{ellsol}) and (\ref{HTq})--(\ref{H32}) in Appendix B.

\subsection{The general case: $\Jg$ and $\Jd$ non-zero.}

\subsubsection{Near the (non--simultaneous) big bang singularity.}

At very early times it is safe to assume that $\HH_q>0$ and $\rho_q>0$ hold. Hence, conditions (\ref{CETcond}) and (\ref{HBcondq}) are equivalent to
\begin{equation} \Drho_q\,\Dh_q\leq 0\quad \Leftrightarrow\quad \dot\sgr\geq 0.\label{condpos}\end{equation}
We have for $t\approx \tbb$ (or $0<a\ll 1$):
\ba \fl \Jg \approx \frac{2(1-\Omega_{q0})\Dig a}{3\Omega_{q0}}\to 0,\quad \Jd\approx \frac{3\Did\HH_{q0}\Omega_{q0}^{1/2}}{a^{3/2}}\to-\infty,\quad \Omega_q\approx 1+O(a)\to 1,\nonumber\label{JgJd1}\\
\ea
where we used (\ref{HTq}) and assumed that $\Did\leq 0$ ($\tbb'\leq 0$) to comply with absence of shell crossings \cite{sussmodes}. These expressions lead to
\begin{equation}\fl  \Drho_q\approx \frac{\Jd}{1-\Jd}\approx -1+\frac{a^{3/2}}{3|\Did|\HH_{q0}\Omega_{q0}^{1/2}},\qquad \Dh_q \approx 2\Drho_q\approx -\frac{1}{2}+O(a^{3/2}), \label{deltasbb} \end{equation}
where we have assumed that $\Did<0$ holds to comply with absence of shell crossings \cite{sussmodes}. Evidently, $\Drho_q$ and $\Dh_q$ have the same sign and so (\ref{condpos}) is violated.

\subsubsection{Asymptotic late times.}

For ever-expanding hyperbolic models we can also assume $\HH_q>0$ and $\rho_q>0$ and test the fulfilment of (\ref{CETcond}) and (\ref{HBcondq}) through (\ref{condpos}). In the asymptotic time range ($t\to\infty$ or $a\to\infty$) we have $\Omega_q\sim O(a^{-1})\to 0$, and from (\ref{HTq})
\ba \fl \Jd\approx \frac{3\Did\sqrt{1-\Omega{q0}}}{a}\to 0,\quad \Jg\approx \Dig\left[1+3\Omega_q\left(1+\ln\left(\frac{\sqrt{\Omega_q}}{2}\right)\right)\right]\to\Dig,\nonumber\label{JgJd2}\\
\ea
hence, from (\ref{Drho}) and (\ref{Dh}), we find that condition (\ref{condpos}) holds:
\begin{equation} \Drho_q\Dh_q \approx \left(\Dig\right)^2\Omega_q\left[1+\ln\left(\frac{\sqrt{\Omega_q}}{2}\right)\right]\leq 0,\end{equation}
since $\Omega_q\ll 1$ and so the logarithmic term inside the square brackets necessarily takes large negative values. The fact that (\ref{condpos}) holds irrespective of the sign of $\Dig$ is an important result, because this sign determines the void ($\Dig\geq 0$) or clump ($\Dig\leq 0$) nature of the time asymptotic radial density profile \cite{sussmodes}.

For elliptic models in their expanding stage ($\HH_q>0$) a late time regime is given by layers approaching the maximal expansion when $\HH_q\to 0$, which corresponds to $t\approx \tmax$ with $t<\tmax$, where $\tmax$ given by (\ref{tmax}), or equivalently $a\to\amax=\Omega_{q0}/(\Omega_{q0}-1)$ and $\Omega_q\sim (a-\amax)^{-1}\to\infty$. From (\ref{HTq}) we have in this limit 
\ba\fl \Jd \approx \frac{3\Did\HH_{q0}(\Omega_{q0}-1)^2}{\Omega_{q0}^{3/2}}(a-\amax)\to 0,\quad \Jg\approx \Dig\left[-2+\frac{3\pi}{2\sqrt{\Omega_q}}\right]\to -2\Dig,\nonumber\label{JgJd3}\\
\ea      
hence, from (\ref{Drho}) and (\ref{Dh}), we find that (\ref{condpos}) holds:
\begin{equation} \Drho_q\Dh_q \approx -\frac{\pi\,(\Dig)^2}{2(1+2\Dig)^2}\sqrt{\Omega_q}\leq 0.\label{condposmax1}\end{equation}
If we approach the maximal expansion limit from the collapsing stage ($\HH_q\to 0$ with $\HH_q<0$), then we need to use (\ref{deltacond}) instead of (\ref{condpos}):
\begin{equation} \rho_q\HH_q\Drho_q\Dh_q=-|\HH_q|\rho_q\Drho_q\Dh_q\leq 0\quad \Rightarrow\quad \Drho_q\Dh_q\geq 0,\label{condposmax2}\end{equation}
with the forms of $\Jg$ and $\Jd$ in the maximal expansion limit obtained from (\ref{HTqc}). This yields the same signs for each one of $\Drho_q$ and $\Dh_q$. Hence, (\ref{condposmax2}) is fulfilled. As a consequence, (\ref{CETcond}) and (\ref{HBcondq}) hold in the maximal expansion limit.

\subsubsection{Collapsing regime.}

As $t\to\tcoll$ we have $\rho_q\to\infty$ and $\HH_q\to-\infty$ as $a\to 0$ and $\Omega_q\to 1$. Hence, conditions (\ref{CETcond}) and (\ref{HBcondq}) are equivalent to (\ref{condposmax2}). From (\ref{HTqc}) and (\ref{H32}) (with $\HH_q<0$) we have in this regime ($\Omega_q-1\approx 0$): 
\begin{equation} \fl \Jg\approx -\frac{3\pi\Dig}{(\Omega_q-1)^{3/2}}\to-\infty,\qquad \Jd\approx -\frac{3\Did(\Omega_{q0}-1)^{3/2}}{\Omega_{q0}(\Omega_q-1)^{3/2}}\to\infty,\label{JgJd4}\end{equation} 
where we assumed that $\Dig\geq 0$ and $\Did\leq 0$ hold to comply with absence of shell crossings \cite{sussmodes}. From the forms above and (\ref{Drho}) and (\ref{Dh}) we obtain
\bse\ba  \Drho_q \approx -1 +\frac{\Omega_{q0}(\Omega_q-1)^{3/2}}{3[\Did(\Omega_{q0}-1)^{3/2}+\pi\Dig\Omega_{q0}]},\label{Drhocoll}\\
 \Dh_q \approx -\frac{1}{2}+\frac{\Omega_{q0}(\Omega_q-1)^{3/2}}{6[\Did(\Omega_{q0}-1)^{3/2}+\pi\Dig\Omega_{q0}]},\label{DHcoll}\ea\ese
which imply that (\ref{condposmax2}) is fulfilled. 

\subsection{Models with suppressed decaying mode.}

From (\ref{Did}), the suppression of the decaying mode, $\Did=0$, is equivalent to a simultaneous Big Bang: $\tbb=\tbbo=$ constant. Following (\ref{tbb}), models of this type (which can be hyperbolic or elliptic) are characterized by initial conditions based on a single free function ($\Omega_{q0}$):
\begin{equation}\HH_{q0}=\frac{Y_{q0}}{t_0-\tbbo},\qquad \Dh_0 = \frac{\Omega_{q0}}{Y_{q0}}\frac{\dd Y_{q0}}{\dd\Omega_{q0}}\DOm_0,\label{ivfSDM}\end{equation}
where $Y_{q0}=Y_q(\Omega_{q0})$ follows from (\ref{Y}) for a given choice of $\Omega_{q0}$ (while $\DOm_0=(r/3)\Omega'_{q0}/\Omega_{q0}$). As shown by the expansions (\ref{JgJd1}), (\ref{JgJd2}), (\ref{JgJd3}) and (\ref{JgJd4}), the decaying mode is only relevant in the very early evolution times $t\approx \tbb$ and near the collapsing singularity (see \cite{sussmodes} for a comprehensive discussion). Therefore, suppression of this mode does not alter the asymptotic forms of perturbations $\Drho_q,\,\Dh_q$, either in the asymptotic time range ($t\to\infty$ of hyperbolic models and $t\to\tmax$ in elliptic models) or near collapse ($t\to\tcoll$). As a consequence,   condition (\ref{condpos}) holds when $\HH_q>0$ or (\ref{condposmax2}) (for $t\to\tmax$ with $t>\tmax$ and $\HH_q<0$), while near collapse the growing mode has the same form as in (\ref{JgJd4}) and the perturbations take the forms (\ref{Drhocoll}) and (\ref{DHcoll}) with $\Did=0$, hence condition (\ref{condposmax2}) is fulfilled. On the other hand, the  suppression of the decaying mode makes an important difference in (\ref{condpos}) in the limit $t\to \tbbo$. The growing mode $\Jg$ takes the same form as in (\ref{JgJd1}), hence the perturbations now take the following forms
\begin{equation} \Drho_q\approx -\frac{2}{5}\Dig(\Omega_q-1)\to 0,\qquad \Dh_q\approx -\frac{1}{3}\Drho_q\to 0,\label{DrhoDh0}\end{equation}
which imply that (\ref{condpos}) is fulfilled as $t\to\tbbo$. This is the opposite result from that when $t\to\tbb$ with a non-zero decaying mode, which indicates a strong relation between early times homogeneity ({\it i.e.} the fact that $\Drho_q,\,\Dh_q\to 0$) and early times positive entropy production (see Figures 1, 3, 5 and 6).

\subsection{Models with suppressed growing mode.}

Since the decaying mode is only dominant at very early times, suppression of the growing mode does not affect the perturbations in the range $t\approx \tbb$, in which (\ref{condpos}) is violated if $\Did\ne 0$. However, the suppression of this mode does affect the late time forms of the perturbations.  Demanding absence of shell crossings \cite{sussmodes}, we have two types of models with a suppressed growing mode:  hyperbolic models complying with $\Dig=0$ and $\HH_q(t-\tbb)>2/3$ (elliptic models with $\Dig=0$ exhibit shell crossings), and parabolic models for which $\Dig\ne 0$ but $\HH_q(t-\tbb)=2/3$. For the hyperbolic models the decaying mode has the same asymptotic time form as in (\ref{JgJd2}), with the perturbations taking the following forms in this limit ($\Omega_q\approx 0$):
\begin{equation}\Drho_q \approx -\frac{3(1-\Omega_{q0})^{3/2}\Did}{\Omega_{q0}}\Omega_q\to 0,\qquad \Dh\approx \frac{1}{3}\Drho_q\to 0,\label{ivfSGM}\end{equation}
which imply that (\ref{condpos}) is violated. For parabolic models ($\KK_q=0$) we have $\HH_q^2=(8\pi/3)\rho_q$, hence we obtain from (\ref{Dadef}): $2\Dh_q=\Drho_q$. Since this is an exact (not asymptotic) relation, then (\ref{condpos}) is violated throughout the whole time evolution and radial domains of parabolic models. 
\begin{figure}
\begin{center}
\includegraphics[scale=0.5]{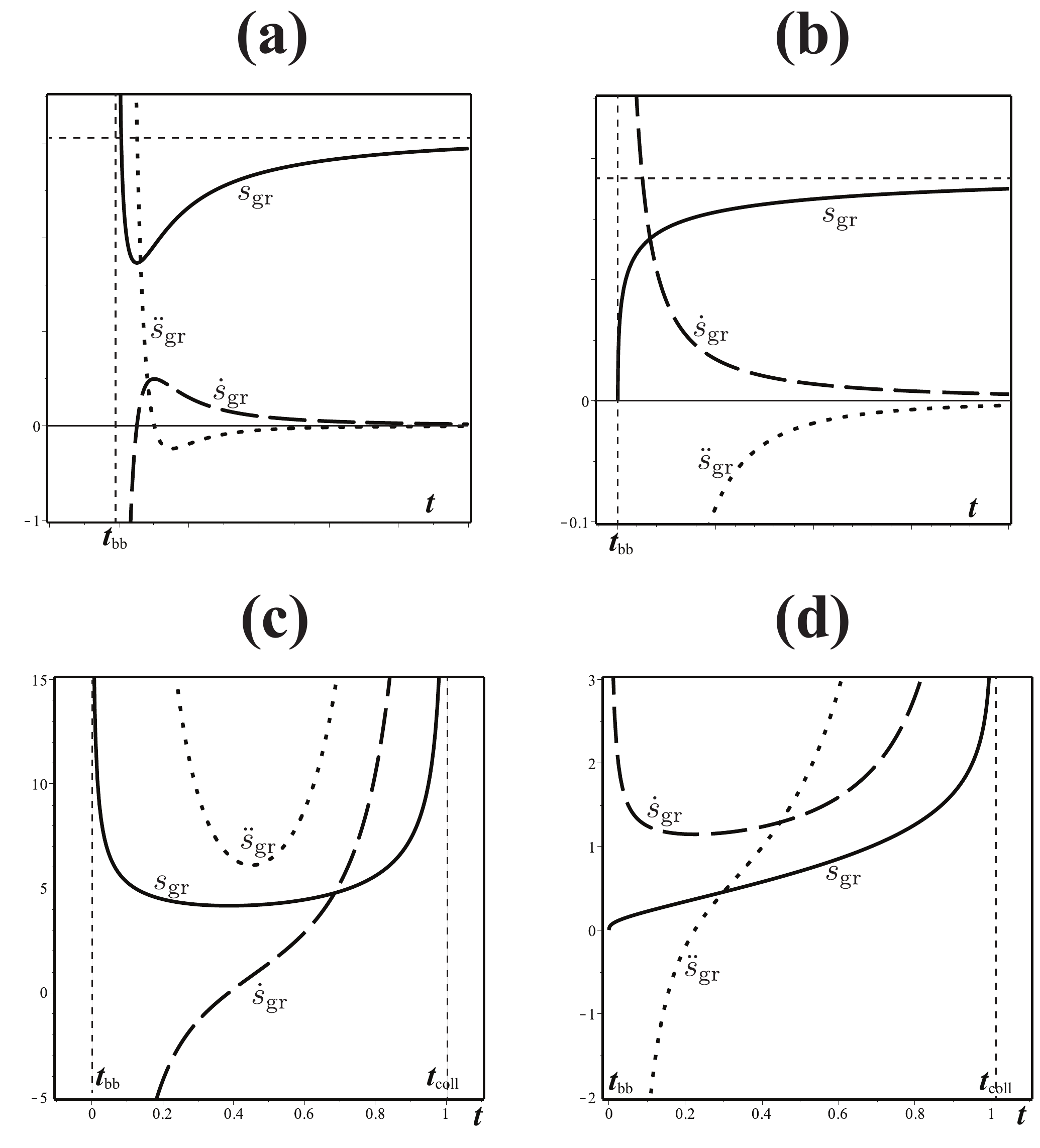}
\caption{{\bf CET entropy time evolution.} The panels display $\sgr,\,\dot\sgr$ and $\ddot\sgr$ as functions of time of a representative dust layer for hyperbolic models (general case in panel (a) and suppressed decaying mode in panel (b)) and elliptic models  (general case in panel (c) and suppressed decaying mode in panel (d)). The forms of the graphs were obtained qualitatively from the results of section 6 and also apply to the HBp and HBq entropies (see detailed discussion in section 7). Notice how the decaying mode (dominant for early times) forces $\sgr$ to diverge as $t\to \tbb$ with $\dot\sgr\to-\infty$, whereas for models with a suppressed decaying mode ((b) and (d)) $\dot\sgr\geq 0$ holds for all times.}
\label{fig1}
\end{center}
\end{figure} 
\section{A qualitative look at the time evolution of the entropies.}
 
We have examined in previous sections the conditions for $\dot\sgr\geq 0,\,\dot\shb{}_p\geq 0$ and $\dot\shb{}_q\geq 0$ and their fulfilment for various asymptotic time ranges. A qualitatively robust description of the full time evolution of these entropies follows by putting together this information. We look at the CET and HB cases separately below (see also Figures 1 and 3). 

\subsection{The CET entropy.}

We obtain directly from (\ref{rhograv2})--(\ref{CETc12}) and (\ref{CETc122}) the exact form of $\dot \sgr$:
\begin{equation} \dot \sgr= -\frac{2\pi\, \alpha\, \rho_{q0}\, \Gamma\, \Dh_q}{|1+3\Dh_q|}\frac{\HH_q\,\Drho_q}{|\HH_q|\,|\Drho_q|},\label{sgrt}\end{equation}
which allows us to examine all relevant sub--cases below:
\begin{itemize} 
\item {\underline {General case}} $\Jg,\,\Jd$ non-zero.  Considering from (\ref{Drho}) that 
\begin{equation} \Gamma =(1+\Drho_{q0})\,(1-\Jg-\Jd),\end{equation}
and using (\ref{JgJd1}) and (\ref{deltasbb}), we have for hyperbolic and elliptic models:
\begin{equation} \fl \dot\sgr\approx \frac{3(1+\Drho_{q0})\Did\HH_{q0}\Omega_{q0}^{1/2}}{a^{3/2}}\to -\infty,\label{CETc12bb}\quad\hbox{as}\quad t\to\tbb,\end{equation}
where we used the fact that $(1+\Drho_{q0})\Did\leq 0$ (to avoid shell crossings) and $\HH_q>0$, so $\HH_q/|\HH_q|=1$ holds in this limit. The limit (\ref{CETc12bb}) implies that $\sgr$ must decrease (for both hyperbolic and elliptic models) from infinite values, but there must always exists an extremum of $\sgr$ since $\dot\sgr$ eventually becomes positive for all models as the evolution proceeds (and $\Jg$ dominates over $\Jd$). The qualitative late time behaviour of $\sgr$ is different for hyperbolic and elliptic models: 
\begin{itemize}
\item Hyperbolic models (see qualitative plot in Figure 1a and numeric plots in Figures 5b and 6). In the asymptotic time regime we have $\HH_q\to 0$ (with $\HH_q>0$) so $\HH_q/|\HH_q|=1$ holds, as well as $\Drho_q\to \Dig/(1-\Dig)$ and $\Dh_q\to 0$ (see also \cite{sussmodes}), then $\dot \sgr\to 0$ holds in the this limit. As a consequence, the extremum of $\sgr$ must be a minimum with $\ddot\sgr>0$, but the curve of $\sgr$ becomes convex as  $\ddot\sgr$ becomes positive with $\dot\sgr\to 0$ and $\sgr$ necessarily reaching a finite (position dependent) asymptotic value. 
\item Elliptic models (see qualitative plot in Figure 1c and numeric plot in Figure 6). In this case $\sgr$ also has an extremum (minimum) as $\dot\sgr$ becomes positive and grows as the maximal expansion is reached (see equations (\ref{condposmax1}) and (\ref{condposmax2})). However, from the fact that $\HH_q<0$ and thus $\HH_q=-|\HH_q|$, the form of $\dot\sgr$ near the collapse singularity has the same form as in (\ref{CETc12bb}) but with a positive sign, which leads to $\dot\sgr\to\infty$ as $t\to\tcoll$. Hence, the form of $\sgr$ as a function of $t$ is convex with $\ddot\sgr>0$ for all the evolution.  
\end{itemize}
\item {\underline {Suppressed decaying mode}}. We have $\dot\sgr\to\infty$ as $t\to\tbbo$ for both hyperbolic and elliptic models, hence $\sgr$ must be finite at $\tbbo$ and display an initial growth with infinite slope. For the late time evolution we have:
 \begin{itemize}
\item Hyperbolic models (see qualitative plot in Figure 1b and numeric plot in Figures 4 and 5a). We have $\dot\sgr\geq 0$ for the full time evolution, but as the evolution proceeds $\dot\sgr$ decreases with $\dot\sgr\to 0$ as $t\to\infty$. As a consequence, for all the evolution $\sgr$ increases towards a the same position dependent terminal value as the general case, but with $\ddot\sgr<0$ for all $t$ (concave curves).
\item Elliptic models (see qualitative plot in figure 1d).  Since  $\dot\sgr> 0$ as $t\to\tmax$, the change of sign of $\dot\sgr$ must occur in the expanding stage, with $\sgr$ reaching a position dependent minimal value for some $t<\tmax$ and growing afterwards with $\dot\sgr\to\infty$ and $\sgr\to\infty$ at the collapse. Hence $\ddot\sgr>0$ holds for the full evolution. 
\end{itemize}
\item {\underline {Suppressed growing mode}}. For both hyperbolic and parabolic models we have $\dot\sgr\to\-\infty$ as $t\to\tbb$ and $\dot\sgr< 0$ with $\dot\sgr\to 0$ as $t\to\infty$, hence $\sgr$ decreases for all the time evolution from infinite to a finite terminal value. 
\end{itemize} 

\subsection{The HBp and HBq entropies.}

In models with a non-zero decaying mode the integrals in (\ref{sHB2}) and (\ref{sHBq}) that define $\shb{}_p$ and $\shb{}_q$, for any domain $\DD[r_b]$, must be evaluated along some time slices that are not complete and everywhere regular: $\rho$ and its average diverge as the slices ``intersect'' the non--simultaneous Big Bang singularity.  The same phenomenon occurs for any domain of elliptic models in which some slices necessarily intersect the non--simultaneous Big Crunch collapsing singularity (see Figure 2). For the regular slices the radial integration range of (\ref{sHB2}) and (\ref{sHBq}) is complete ({\it i.e.} $0\leq r\leq r_b$), but for slices hitting singularities it is restricted by the conditions $\tbb'\leq 0$ and $\tcoll'\geq 0$ that follow from demanding absence of shell crossings. For any domain bounded by a finite $r=r_b$, the effects of the non--simultaneity of the singularities on (\ref{propvol})--(\ref{Mp}) and (\ref{rhoaveq2})--(\ref{sHBq}) are illustrated by Figure 2 and summarized below:
\begin{itemize}
\item Near the Big Bang in hyperbolic and elliptic models or regions (see Figure 2a): for time slices $\tbb(0)>t_s>t_{(-)}=\tbb(r_b)$ we have $\rho\to \infty$ and $\HH\to \infty$ and $a\to 0$ as $r\to r_s$. Hence, all integrals in (\ref{sHB2}) and (\ref{sHBq}) and in (\ref{Sdot1}) and (\ref{Sdot1q}) are improper integrals whose convergence in this limit must be verified in the restricted integration range $r_s<r\leq r_b$, where $t_s=\tbb(r_s)$. As the slices approach $t=t_{(-)}$, the spherical domains $\DD[r_b]$ ``shrink'' to very thin shells around the singular point $[r_s,\tbb(r_s)]$ corresponding to a ``narrowing'' of the radial range as $r_b\approx r_s$. The limit $t_s\to t_{(-)}$ is equivalent to both, the convergence limit $r\to r_s$ and the limit $r_s\to r_b$ of ``narrow'' domains (see Figure 2a). As proven in Appendix D.2, for every fixed arbitrary $r_b$ we have
\bse\ba \fl \lim_{r_s \to r_b} \shb{}_p[r_b]-\shb{}_p^{(\textrm{\tiny{eq}})}=0,\qquad \lim_{r_s \to r_b} \shb{}_q[r_b]-\shb{}_q^{(\textrm{\tiny{eq}})}=0,\label{shblim}\\
\fl \lim_{r_s \to r_b} \dot\shb{}_p[r_b]=\infty,\qquad  \lim_{r_s \to r_b} \dot\shb{}_q[r_b]=\infty,\label{dotshblim}\ea\ese
which imply that the very early time behaviour of $\shb{}_p$ and $\shb{}_p$ is radically different from that of $\sgr$:  for every domain (since $r_b$ is arbitrary) their earliest time value at some $t=t_{(-)}$ is finite and given by the ``equilibrium'' entropies (these can be a different integration constant for different domains), with the initially instantaneous infinite entropy growth depicted by Figures 3a and 3c for $t$ close to $t=t_{(-)}$ (this behaviour is analogous to the very early time evolution of $\sgr$ depicted by Figures 1b and 1d for models with a suppressed decaying mode).
\item Near the Big Crunch in elliptic models or regions (see Figures 2b, 3c and 3d): in slices $\tcoll(0)<t_s<t_{(+)}=\tcoll(r_b)$  the integration range is also $r_s<r\leq r_b$, but now with $t_s=\tcoll(r_s)$ and with the lower bound $r_s$ of the integral in (\ref{sHB2}) also approaching $r_b$ as $t_s\to t_{(+)}$. The limit (\ref{shblim}) also holds for $\shb{}_p[r_b]$ and $\shb{}_q[r_b]$, but not (\ref{dotshblim}). Instead, we have (see Appendix D.2): 
\begin{equation} \lim_{r_s \to r_b} \dot\shb{}_p[r_b]=-\infty,\qquad  \lim_{r_s \to r_b} \dot\shb{}_q[r_b]=-\infty,\label{dotshblimc}\end{equation}
As a consequence, the values of $\shb{}_p[r_b]$ and $\shb{}_q[r_b]$ plummet down with a final infinite slope to their final equilibrium value (the same as in the Big Bang) for every domain $r_b$.
\end{itemize} 
Because of (\ref{dotshblim}), the values of $\shb{}_p$ and $\shb{}_q$ must increase from their equilibrium values at least for domains bounded by $r_b$ that intersect the non--simultaneous Big Bang in slices $t_s$ close to $t_{(-)}$.
While for every domain the terminal collapse value of the HB entropies is the same as their initial Big Bang value in elliptic models, the terminal time asymptotic values as $t\to\infty$ in hyperbolic models is necessarily larger than their initial Big Bang values at $t=t_{(-)}$ (see proof in Appendix D.3).

This information on the early and late asymptotic time limits, together with the sufficient conditions for the production of entropy, are sufficient to obtain a qualitative picture of the time evolution of $\shb{}_p[r_b]$ and $\shb{}_q[r_b]$ when both density modes are nonzero (this is depicted by Figures 3a and 3c). The CET and HB entropies clearly exhibit a different behaviour for slices intersecting non--simultaneous singularities, as the integration domains of the integrals that define $\dot\shb{}_p$ and $\shb{}_q$ in (\ref{HBcond}) and (\ref{HBcondq}) are incomplete (this follows from (\ref{shblim})--(\ref{dotshblim}) and (\ref{dotshblimc})). However, for slices $t>\tbb(0)$ (see figure 2a) and $\tbb(0)<t<\tcoll(0)$ (elliptic models, figure 2b) these integrals are evaluated for arbitrary complete domains $0\leq r\leq r_b$. Since their integrands are similar to those of the CET entropy in (\ref{CETcond}), the HBp and HBq entropies should exhibit for these complete slices a qualitatively similar evolution (see Figures 3a and 3c) to that displayed by Figures 1a and 1c for the CET entropy. This assessment follows from the fact that the fulfillment of $\dot\sgr( r)\geq 0$ for every $0\leq r\leq r_b$ implies for complete slices (through the sufficient conditions (\ref{HBcond1})--(\ref{HBcond2}) and (\ref{HBcond3})) the fulfillment of $\dot\shb{}_p\geq 0$ and $\dot\shb{}_q\geq 0$ for $r=r_b$. Moreover, these sufficient conditions may place strong restrictions on the radial profiles of $\rho$ and $\HH$, therefore we cannot exclude the possibility that well behaved models exist in which $\dot\shb{}_p< 0$ and/or $\dot\shb{}_q<0$ could hold in some subdomains $r<r_b$ of $\DD[r_b]$, or in restricted time ranges of complete slices, even if $\dot\sgr(r_b)>0$ holds for these times.     

The evaluation (and early time evolution) of $\shb{}_p$ and $\shb{}_q$ is completely different for models (hyperbolic and elliptic) with a suppressed decaying mode, as in this case the Big Bang is simultaneous ($\tbb=\tbbo$), hence all slices $t>\tbbo$ are regular and the radial integration range of $\shb{}_p$ and $\shb{}_q$ is complete for all domains (there is no need to consider improper integrals). As we prove in Appendix D.4, the integrals in the right hand sides of (\ref{sHB2}) and (\ref{sHBq}) vanish  and $\dot\shb{}_p\to\infty,\,\dot\shb{}_q\to\infty$ in the limit $t\to\tbbo$, which are the same limits (\ref{shblim})--(\ref{dotshblim}) but valid for all domains. Hence, as depicted by figures 3b and 3d, $\shb{}_p$ and $\shb{}_q$ increase from their initial equilibrium value at $t=\tbbo$ for all domains and follow similar evolution patterns as those displayed in Figures 1b and 1d (though the behaviour of $\shb{}_p$ and $\shb{}_q$ near the Big Crunch is the same as in the general case: compare the curves of Figures 3c and 3d as $t\to t_{(+)}$).      
\begin{figure}
\begin{center}
\includegraphics[scale=0.4]{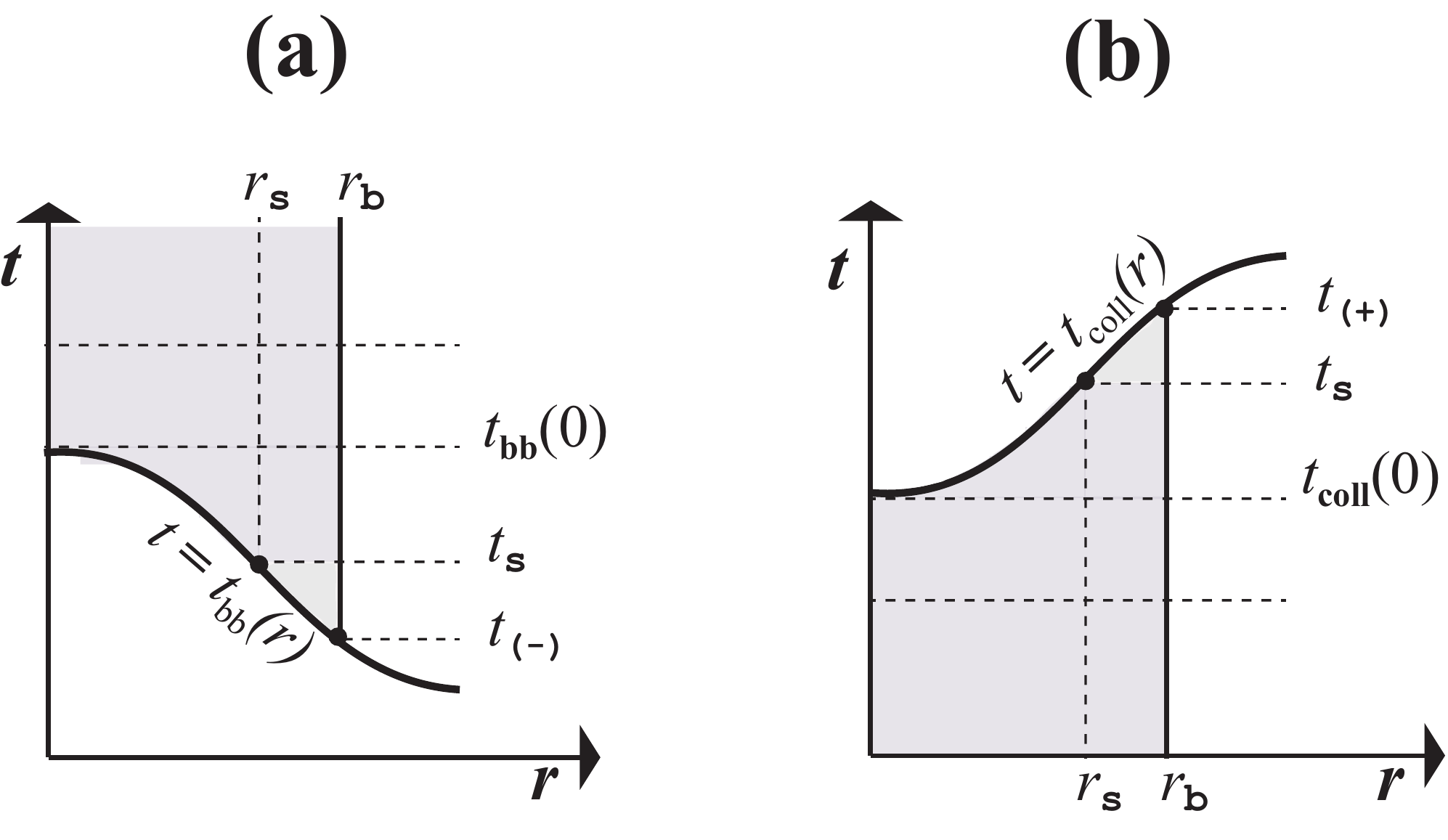}
\caption{{\bf HB entropies near non--simultaneous singularities.} The panels display the time evolution of a generic domain bounded by $r=r_b$ (shaded region) in time slices close to non--simultaneous singularities: the Big Bang ($t=\tbb( r)$ in panel (a)) and the collapsing Big Crunch ($t=\tcoll( r)$ in panel (b)). Notice that for a typical time slice $t=t_s$ that ``intersects'' these singularities $\rho$ and $\HH$ diverge as $r\to r_s$. Therefore, the integrals in (\ref{sHB2}) and (\ref{sHBq}) for any domain $r=r_b$ must be evaluated in the restricted range to $r_s<r\leq r_b$. Since these integrals are improper, we need to verify their convergence as $r\to r_s$ (see Appendix D.2).}
\end{center}
\end{figure} 
\begin{figure}
\begin{center}
\includegraphics[scale=0.5]{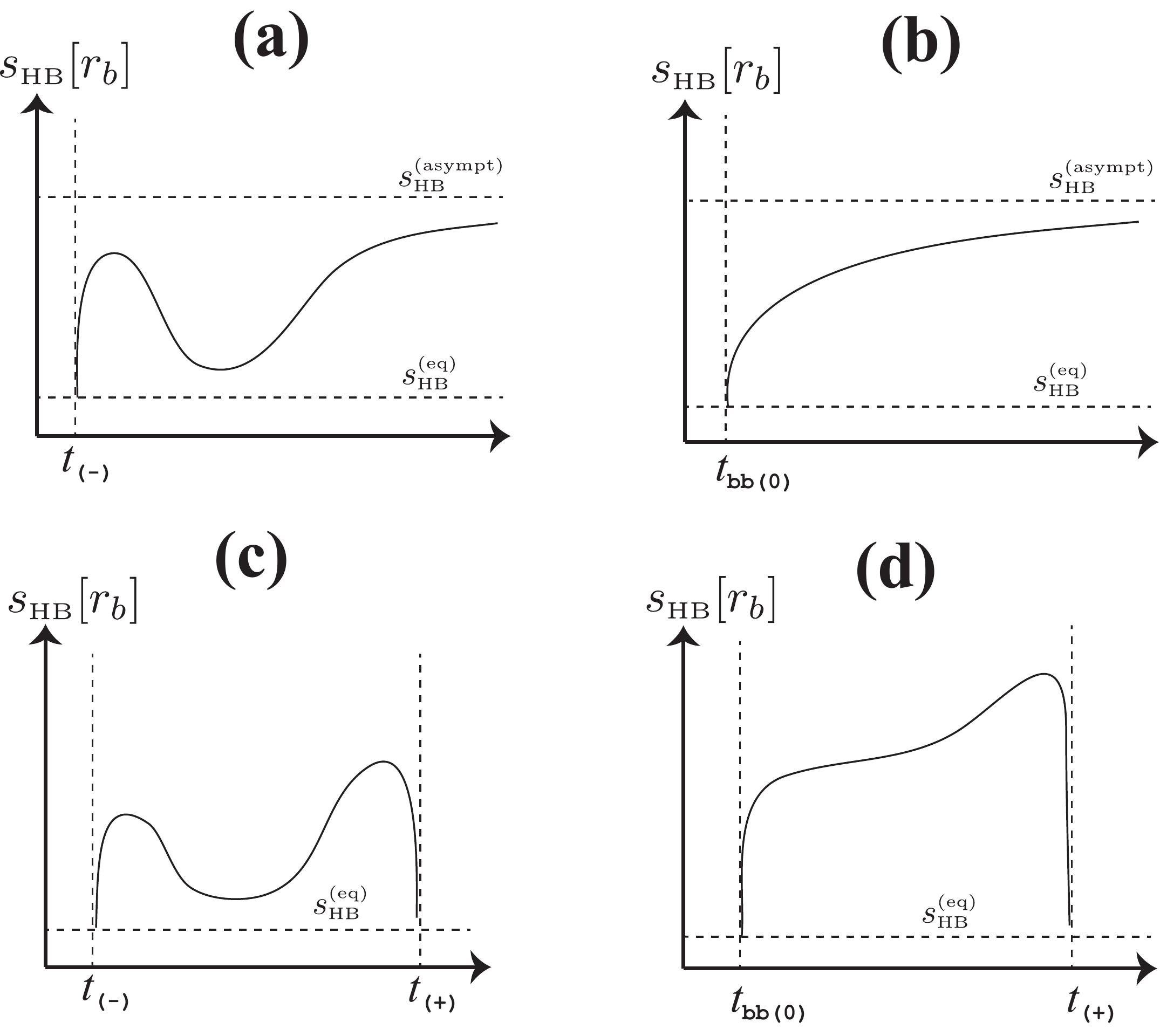}
\caption{{\bf Time evolution of the HB entropies.} The panels display $\shb[r_b]$ (which can be either one of $\shb{}_p[r_b]$ or $\shb{}_q[r_b]$) as a function of time for a representative generic domain $\DD[r_b]$ with $r_b$ finite. Hyperbolic and elliptic models respectively correspond to panels (a)--(b) and ( c)--(d). Panels (a) and ( c) depict  the cases with non-zero decaying mode, while the cases with zero decaying mode (simultaneous Big Bang $t=\tbbo$) are depicted by panels (b) and (d). The forms of the graphs were obtained qualitatively from the discussion in section 7.2 and the proofs of convergence of the two HB entropies in Appendix D.2 and D.4 (see the text of these sections for further detail).}
\end{center}
\end{figure} 
\section{Numerical examples.}

We complement the qualitative study of the previous section by the following three numeric examples:
\footnote{We only examine the CET entropy. The second and third examples are meant to illustrate the behaviour of entropy production $\dot\sgr$. They are not meant to be ``realistic'' or to comply with observational constraints.}   

\subsection{Cosmological void with suppressed decaying mode.}

Figure 4 displays $\log |\Del_q(\rho)\Del_q(\HH)|$ as a function of $t$ and $r$ for the LTB void model studied in \cite{February:2009pv} (model \# 2), which is radially asymptotic to an Einstein--de Sitter FLRW model. The  free parameter is a present day ($a=a_0=1$) matter density profile (denoted by `` $\Omega_m$'' in \cite{February:2009pv}) that exactly corresponds to the initial value function:
\begin{equation}\Omega_{q0}( r)=\Omega_{\textrm{\tiny{out}}}-\left(\Omega_{\textrm{\tiny{out}}}-\Omega_{\textrm{\tiny{in}}}\right)\exp\left(-\frac{r^{2}}{\sigma^{2}}\right),\label{OmSDM}\end{equation}
where the parameters were selected from compliance with best-fit values to SN1a and age data (more details in \cite{February:2009pv}): $H_{0}=64.33\mbox{ km s}^{-1}\mbox{Mpc}^{-1}$, $\Omega_{\textrm{\tiny{in}}}=0.120$, $\Omega_{\textrm{\tiny{out}}}=1$ and $\sigma=3.77\mbox{ Mpc}$, with cosmic age $t_0 =13.46$ Gyr (the remaining initial value functions $\HH_{q0}, \Dh_{q0},\,\DOm_{q0}$ follow from (\ref{ivfSDM}) with $\tbbo=0$). We see that condition (\ref{CETcond}) is fulfilled: $\Del_q(\rho)\Del_q(\HH)<0$ holds all across spacetime, thus showing that entropy production is positive for the whole time evolution of all observers.

\begin{figure}
\begin{center}
\includegraphics[scale=0.4]{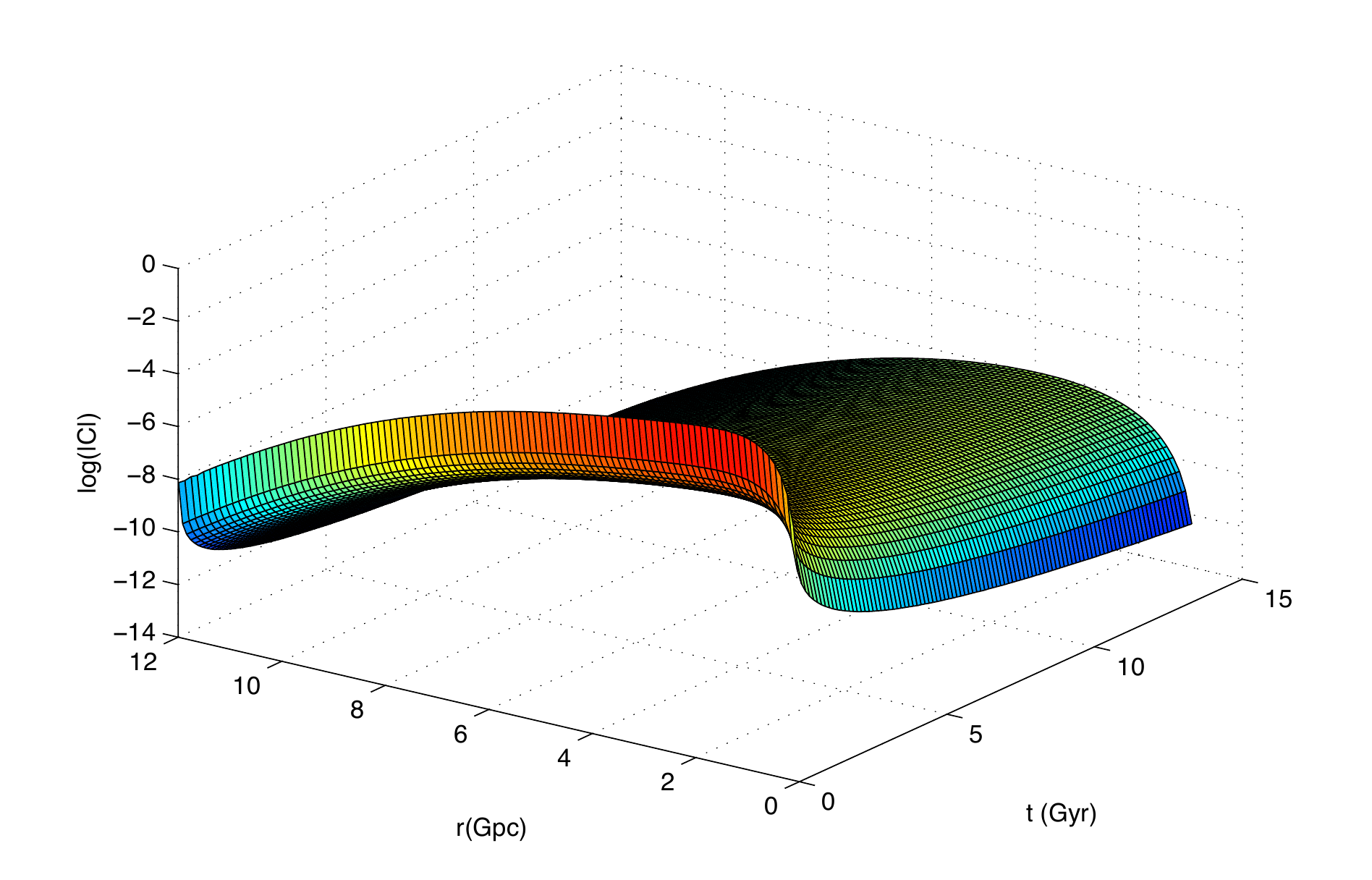}
\caption{$\ln |\Del_q(\rho)\Del_q(\HH)|$ as a function of $t$ and $r$ for model \# 2 of \cite{February:2009pv} described in the text. $\Del_q(\rho)\Del_q(\HH)$ is negative throughout spacetime, which implies a positive entropy production everywhere in spacetime.} 
\end{center}
\end{figure}

\subsection{Cosmological void model with non-zero decaying mode.}

As a second example we consider a hyperbolic void model whose decaying mode is non-zero, but otherwise it is almost identical to the previous one: it is also asymptotic (in the radial direction) to an Einstein--de Sitter background. This model follows from the same form of $\Omega_{q0}( r)$ in (\ref{OmSDM}), with its second initial value function $\HH_{q0}( r)$ not given by (\ref{ivfSDM}), but by
\begin{equation}\fl  \HH_{q0}=\frac{Y_{q0}(\Omega_{q0})}{t_0-\tbb( r)},\qquad \tbb=0.01\left[\exp\left(-\frac{r^2}{\sigma_0^2}\right)-1\right],\qquad \sigma_0=1\,\hbox{Mpc},\label{OmDM} \end{equation}
which, evidently, introduces a small decaying mode via a non--simultaneity of the Big Bang (marked $\tbb=\tbbo=0$ in the previous example). This yields a position dependent cosmic age that goes from $t_0=13.46$ Gyr for central observers to an asymptotic value 1\% larger ($\sim 10^8$ years) for observers in the Einstein--de Sitter background. We plot in Figure 5 the product $-\Drho_q\Dh_q$ obtained from the numerical solution of the system (21a)--(21d) in \cite{part2}, which is proportional to the sign of $\dot\sgr$ (from (\ref{condpos})). Comparison of panels (a) and (b) of this figure reveals that $\dot\sgr>0$ behaves as depicted in the qualitative plots in Figures 1a and 1b: it has almost identical form for both models, save for early very times $a<10^{-3}$ in which $\dot\sgr$ in panel (b) becomes negative and tends to $-\infty$ as $t\to\tbb$ (or $a\to 0$). Since the present day time is taken as $a=a_0=1$, these early times can be identified with times before the last scattering surface $z\sim 1000$ where the dust source is no longer a valid model of cosmic matter.      

\begin{figure}
\begin{center}
\includegraphics[scale=0.4]{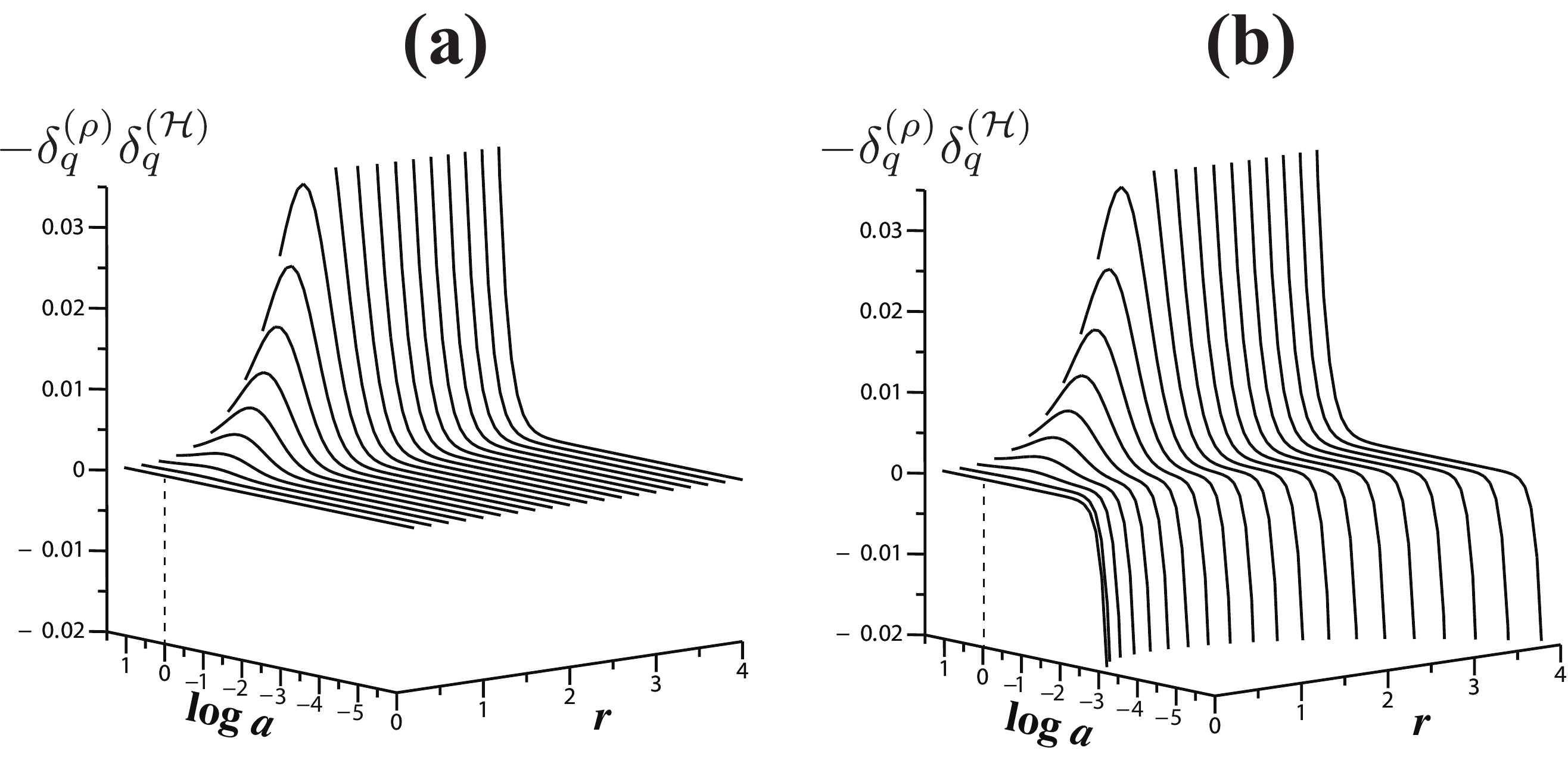}
\caption{{\bf Entropy production at early times in a void model with zero and non-zero decaying mode.} The figure depicts  the plot of $-\Drho_q\Dh_q \propto \dot\sgr$ as a function of $\log a$ and $r$ marking the present day time as $a=a_0=1$. Panel (a) corresponds to the model of Figure 2 with suppressed decaying mode and panel (b) corresponds to the closely related model with a non-zero decaying mode whose initial value functions are (\ref{OmSDM}) and (\ref{OmDM}). Notice that the curves have the forms of $\dot\sgr$ depicted qualitatively in figures 1a and 1b and that the effects of the decaying mode (change of sign of $\dot\sgr$) are only significant for very early times $a< 10^{-3}$ in panel (b). }
\label{fig3}
\end{center}
\end{figure} 

\subsection{Dust gravitational collapse.}

The third example is furnished by a ``spherical collapse model'' defined by a ``mixed'' elliptic/hyperbolic configuration, so that  ``inner'' dust layers near the symmetry centre (elliptic region) collapse to a non--simultaneous Big Crunch, while ``external'' layers perpetually expand (hyperbolic region). The initial value functions are given by
\begin{equation}\fl  \frac{4\pi\rho_{q0}}{3H_0^2}=\frac{\Omega_0\,(1+\epsilon_1 + x^2)}{2(1+x^2)},\qquad \frac{\KK_{q0}}{H_0^2}=\frac{(\Omega_0-1)(1+\epsilon_2+x^{3/2})}{1+x^{3/2}},\end{equation}
with the same value of $H_0$ as in the first example, $\Omega_0=0.8,\,\epsilon_1=2.0,\,\epsilon_2=-1.25,\,\,x=r/\sigma_0$ with $\sigma_0$ an arbitrary length scale and $t$ is normalized with the Hubble factor at the last scattering surface $t_{\textrm{\tiny{LS}}}$ (so that present day is $t_0\sim 10^4$). We have a collapsing elliptic region ($\KK_{q0}>0$) for $0\leq x<0.397$ and an expanding hyperbolic region ($\KK_{q0}<0$) for $r>0.397$. The curves of Figure 6, which were obtained by solving numerically the system of evolution equations (\ref{EVq21})--(\ref{EVq24}), clearly reveal how $\dot\sgr\to\infty$ as $t\to\tcoll$ in the collapsing layers of the elliptic region (like the late time form of $\dot\sgr$ in Figure 1c), whereas $\dot\sgr\to 0$ as $t\to\infty$ in the expanding layers of the hyperbolic region (as in its late time form in figure 1a). The early time evolution is not displayed by Figure 6, but since $\Jd\ne 0$ it is qualitatively analogous to that depicted by figures 1a, 1c and 5b ($\dot\sgr\to-\infty$ as $t\to\tbb$) .

\begin{figure}
\begin{center}
\includegraphics[scale=0.4]{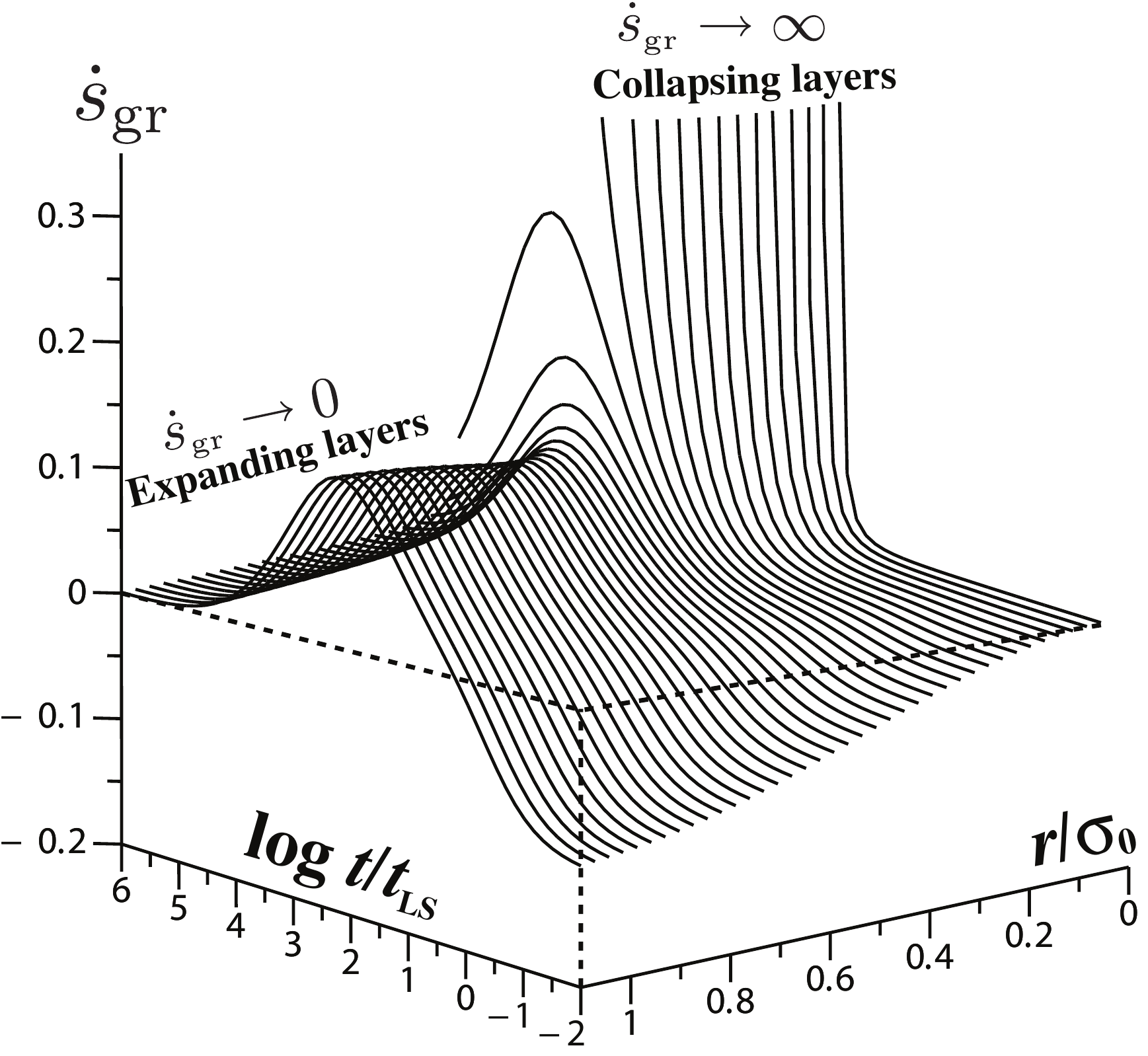}
\caption{{\bf Entropy production in a collapse/expansion evolution.} The figure displays $\dot\sgr$ given by (\ref{sgrt}) as a function of $\log t$ and $r$ for a mixed configuration made of an ``inner'' elliptic region surrounded by an ``outer'' hyperbolic region that expands perpetually. Notice that $\dot\sgr\to\infty$ for collapsing layers while $\dot\sgr\to 0$ for asymptotic layers as $t\to\infty$. }
\label{fig1}
\end{center}
\end{figure} 

\section{Radial scaling and asymptotic behaviour.}    

\subsection{Integrability conditions for the CET entropy.} 

As opposed to the HBp and HBq entropies, the CET proposal defines entropy through the convective derivative $\dot\sgr$ in the Gibbs equation (\ref{CET2}), which is a relation between entropy and energy one--forms with integrating factor $\Tgr$:
\begin{equation}\dd \sgr =\frac{\dd (\rhogr V)}{\Tgr}.\end{equation}
For the spherically symmetric LTB models in  coordinates $(t,r,\vartheta,\phi)$ we have $\dd \sgr=[\dot\sgr,\,\sgr',0,0]$, and so the components of the Gibbs one form yield an integrability condition
\begin{equation}\fl \dot \sgr =\frac{(\rhogr V)\,\dot{}}{\Tgr},\quad \sgr' =\frac{(\rhogr V)'}{\Tgr} \quad \Rightarrow\quad \Tgr'\dot\sgr - \dot\Tgr\sgr'=0,\end{equation}
which leads to the following fully general expression for $\sgr'$: 
\begin{equation}    \sgr' =-\frac{\partial_r[F(\Tgr)]}{\partial_t[F(\Tgr)]}\,\dot\sgr,\label{sgrrdef}\end{equation}
where $F(\Tgr)$ is an arbitrary smooth function. Considering for simplicity the particular case $F(\Tgr)=\Tgr^2$, and substituting the forms of $\Tgr$ and $\dot \sgr$ in (\ref{Tgrav2}), (\ref{CETc12}), (\ref{CETc122}) and (\ref{sgrt}) into (\ref{sgrrdef}), yields after some algebraic manipulation 
\begin{equation} \fl \sgr' = -\frac{3\alpha\,\Omega_q\,a^3}{4\,H_{q0}^2}\,\frac{\Gamma\,\Dh_q\,\Drho_q\,\HH_q}{|1+3\Dh_q|\,|\Drho_q|\,|\HH_q|}\,\frac{\HH'_q\,(1+3\Dh_q)+3\HH_q(\Dh_q)'}{[1+3\Dh_q+\frac{1}{2}\Omega_q\,(1+3\Drho_q)]},\label{sgrr}\end{equation}
where $\Omega_q$ is given by (\ref{OmDOm}) and we eliminated $2\pi\rho_{q0}/\HH_q^2,\,\dot\HH_q,\,\dot\Dh_q$ from (\ref{HHq}) and the evolution equations (19b) and (19d) of \cite{part2}. 

\subsection{Radial asymptotic convergence of LTB models.} 

In order to examine the radial asymptotic behaviour of the CET and the two HB entropies we will rely on the results of the comprehensive study of the radial asymptotics of LTB models undertaken in \cite{RadAs}. Assuming absence of shell crossing singularities and a well behaved radial coordinate (since radial rays are spacelike geodesics, the proper radial length along them must be a monotonous function of $r$), we characterize the radial asymptotic behaviour of the covariant parameters of the models by means of the  definition of radial asymptotic convergence ``$A\sim \tilde A$\,'' given in section 7.1 of \cite{RadAs}, considering the polynomial asymptotic forms for the following initial value functions:
\begin{equation} \fl \rho_{q0}\sim m_0+m_1r^{-\alpha},\qquad \KK_{q0}\sim k_0+k_1 r^{-\beta},\qquad \HH_{q0}\sim H_0+H_1\,r^{-\nu},\label{asrhok}\end{equation}
where $m_0\geq 0,\,m_1,\,k_0,\,k_1,\,H_0,\,H_1,\,\alpha,\,\beta$ and $\nu$ are real constants, whose values and restrictions (given in Table 1)  correspond to the various classes of radial asymptotic convergence studied in \cite{RadAs}. Using (\ref{asrhok}) yields the following asymptotic forms for the scale factors:
\begin{equation} \\
 a\sim 1+ (H_0+H_1 r^{-\nu})(t-t_0),\qquad \Gamma \sim 1-\frac{\nu H_1 r^{-\nu}\,(t-t_0)}{1+H_0 (t-t_0)},\label{asaG}\end{equation}
obtained by looking at the radial asymptotic behaviour of the exact solutions (\ref{hypsol})--(\ref{ellsol}) (expressed in terms of $a,\,\rho_{q0}$ and $\KK_{q0}$) for the various classes of radial asymptotic convergence listed in Table 1.

From (\ref{asrhok}) and (\ref{asaG}) and considering the values of the asymptotic parameters given in Table 1, the asymptotic radial limits of the q--scalars and their perturbations for the various classes of models are given below (see \cite{RadAs} for a comprehensive discussion):
\begin{itemize}
\item Models asymptotic to FLRW: Einstein de Sitter, open FLRW and Milne ($H_0>0$)
\begin{equation} \fl (a,\,\rho_q,\,\HH_q,\,\Omega_q)\to (\tilde a,\,\tilde\rho,\,\tilde\HH,\,\tilde\Omega),\quad (\Drho_q,\,\Dh_q)\to 0,\quad \Gamma\to 1,\label{asympt1}\end{equation}
where $\tilde a,\,\tilde\rho,\,\tilde\HH,\,\tilde\Omega$ are the scale factor, density, Hubble scalar and Omega factor of the FLRW model ($\tilde\rho=0$ for models asymptotic to Milne).
\item Models asymptotic to Minkowski: MD, VD and G ($H_0=0$, see Table 1)
\bse\ba \fl (\rho_q,\,\HH_q)\to 0,\quad (\Drho_q,\,\Dh_q)\to (\Drho_{q_\infty},\,\Dh_{q_\infty}),\quad (a,\,\Gamma)\to 1,\label{asympt2}\\
\fl \Omega_q\to 0\,\, \hbox{(VD)},\qquad \Omega_q\to 1\,\, \hbox{(MD)},\qquad \Omega_q\to \frac{8\pi m_1}{8\pi m_1+3k_1}<1\,\, \hbox{(G)},\label{asympt3}\ea\ese
where $\Drho_{q_\infty}=-\alpha/3$ and $\Dh_{q_\infty}=-\nu/3$ are the finite non-zero asymptotic values of the perturbations. 
\end{itemize}

\subsection{The CET entropy.}

It is evident from (\ref{sgrt}), (\ref{asympt1}) and (\ref{asympt2}) that $\dot\sgr\to 0$ as $r\to\infty$ for all convergence classes, as for models converging to FLRW we have in this limit $\rho_{q0}\to m_0>0$ but $\Dh\to 0$, while for models converging to Milne or Minkowski we have $\Dh\to -\nu/3<0$ but $\rho_{q0}\to 0$.

Regarding $\sgr'$ in (\ref{sgrr}), the asymptotic forms for $\Omega_q,\,a$ and $\Gamma$ in (\ref{asympt1}) and  (\ref{asympt2})--(\ref{asympt3}) are non-zero for all models (save $\Omega_q$  for the VD hyperbolic models). However,  we have for all convergence classes the following asymptotic form
\begin{equation}\fl  \HH'_q (1+3\Dh_q) + 3\HH_q (\Dh_q)' \sim \frac{\nu(\nu-1)\,H_1\,r^{-\nu-1}}{[1+H_0(t-t_0)]^2}+O(r^{-2\nu-1})\to 0,\end{equation}
where we used (\ref{Dadef}), (\ref{HHq}) and the asymptotic forms (\ref{asaG}). Therefore, we have $\sgr'\to 0$ for all models compatible with a radial asymptotic range. Since both $\dot\sgr$ and $\sgr'$ tend to zero as $r\to\infty$, then $\sgr$ (which is defined up to an additive constant) must reach a finite constant asymptotic ``equilibrium'' value in the radial direction for all time slices.  

%
\begin{table}
\begin{center}
\begin{tabular}{|c| c| c|}
\hline
\hline
\hline
\hline
\multicolumn{3}{|c|}{{\bf Parameters of radial asymptotic convergence.}}
\\
\hline
\hline
\hline
\hline
\multicolumn{3}{|c|}{Hyperbolic models, $\KK_{q0}<0$ or $0<\Omega_{q0}<1$.}
\\  
\hline
\hline
\hline
\hline
{Asymptotic class} &{$m_0,\,k_0,\,\alpha,\,\beta$} &{$H_0,\,\nu$} 
\\
\hline
{{\bf open FLRW}} &{$m_0>0,\,k_0<0$} &{$H_0=[2m_0+|k_0|]^{1/2}$ }
\\  
{} &{$\alpha>0,\,\beta>0 $} &{$\nu =\hbox{min}(\alpha,\beta)$}
\\
\hline  
{{\bf Milne}} &{$m_0=0,\,k_0<0$} &{$H_0=|k_0|^{1/2}$}
\\  
{} &{$0<\alpha\leq 3,\,\,\beta>0$} &{$\nu =\hbox{min}(\alpha,\beta)$} 
\\ 
\hline
{{\bf Einstein de Sitter}} &{$m_0>0,\,k_0=0$} &{$H_0=[2m_0]^{1/2}$}
\\  
{} &{$\alpha>0,\,\,0<\beta\leq 2$} &{$\nu =\hbox{min}(\alpha,\beta)$}
\\
\hline
{{\bf Minkowski}} &{$m_0=k_0=0$} &{$H_0=0,\,H_1>0$}
\\  
{} &{$0<\alpha\leq 3,\,\,0<\beta\leq 2$} &{$\nu =\alpha/2$ (MD: $\beta>\alpha$)} 
\\
{} &{} &{$\nu =\beta/2$ (VD: $\beta<\alpha$)}
\\
{} &{} &{$\nu =\gamma/2$ (G: $\gamma=\beta=\alpha$)}
\\
\hline
\hline
\hline
\hline
\multicolumn{3}{|c|}{Elliptic models, $\KK_{q0}>0$ or $\Omega_{q0}>1$. }
\\
\hline
\hline
\hline
\hline
{Asymptotic class} &{$m_0,\,k_0,\,\alpha,\,\beta$} &{$H_0,\,\nu$} 
\\
\hline
{{\bf Einstein de Sitter}} &{$m_0>0,\,k_0=0$} &{$H_0=[2m_0]^{1/2}$}
\\  
{} &{$\alpha>0,\,\,\beta\geq 2$} &{$\nu =\hbox{min}(\alpha,\beta)$}
\\    
\hline
{{\bf Minkowski}} &{$m_0=k_0=0$} &{$H_0=0,\,H_1>0$}
\\   
{} &{$0<\alpha\leq 3,\,\,\beta\geq 2$} &{$\nu =\alpha/2$ (MD $\beta>\alpha$)}
\\   
\hline
\hline
\hline
\hline
\end{tabular}
\end{center}
\caption{{\bf{Parameters in the classification of radial asymptotic convergence}}. The values for the parameters describe all the classes of radial asymptotic convergence of LTB models discussed in \cite{RadAs} for polynomial trial functions (\ref{asrhok}). Models converging to Minkowski are subdivided in the following classes according to the limit of $\alpha_{q0}\propto \KK_{q0}/\rho_{q0}$ as $r\to\infty$: ``MD'' (matter dominated, $\alpha<\beta$) if $\alpha_{q0}\to 0$, ``VD'' (vacuum or curvature dominated, $\alpha>\beta$) if $\alpha_{q0}\to \infty$ and ``G'' (generic, $\alpha=\beta$) if $\alpha_{q0}\to \alpha_0=$ constant. The restrictions on the range of $\alpha$ and $\beta$ in models converging to Milne, Einstein de Sitter and Minkowski strictly follow from the conditions to avoid shell crossings examined in \cite{RadAs}. The relation between $\alpha,\,\beta$ and $\nu$ follows from the asymptotic behavior of $\HH_{q0}^2=(8\pi/3)\rho_{q0}-\KK_{q0}$ for the different classes of convergence. The values $\alpha=3$ and $\alpha=\beta=2$ respectively correspond to models asymptotic to Schwarzschild and to self similar dust solutions (see further detail in \cite{RadAs}).}
\label{tabla3}
\end{table}

\subsection{The HBp and HBq entropies.}

The radial scaling of these entropies depends on the behaviour of the integrals (\ref{sHB2}) and (\ref{sHBq}) as functions of an increasing domain boundary $r_b$, up the asymptotic limit $r_b\to\infty$ that would correspond to domains that encompass whole time slices ($t$ constant hypersurfaces). While (\ref{sHB2}) and (\ref{sHBq}) are analogous to the proper volume mass--energy integrals $M_p$ and $M_q$ in (\ref{Mp}) and (\ref{Mq}), with $\rho\ln[\rho/\rhoav_p]$ and $\rho\ln[\rho/\rhoav_q]$ playing the role of entropy ``densities'', the asymptotic convergence of these ``densities'' does not imply the asymptotic convergence of the proper volume integrals (\ref{sHB2}) and (\ref{sHBq}), just as a converging $\rho$ does not prevent the mass--energy functionals $M_p$ and $M_q$ in (\ref{Mp}) and (\ref{Mq}) from diverging in the radial asymptotic range. 

In order to explore the asymptotic convergence of $\shb{}_p$ and $\shb{}_q$ we use  (\ref{propvol}) and (\ref{qvol}) to rewrite (\ref{sHB2}) and (\ref{sHBq}) as    
\bse\ba \fl \shb{}_p[r_b] -\shb{}_p^{(\textrm{\tiny{eq}})}\nonumber\\ 
\fl = 4\pi\gamma_0\left[\int_0^{r_b}{\frac{\rho_{q0}(1+\Drho_0)\,\bar r^2}{\sqrt{1-\KK_{q0} \bar r^2}}\ln\rho\,\dd \bar r}-\ln\rhoav_p[r_b]\int_0^{r_b}{\frac{\rho_{q0}(1+\Drho_0)\,\bar r^2}{\sqrt{1-\KK_{q0} \bar r^2}}\dd \bar r}\right],\label{sHBpp}\\
\fl \shb{}_q[r_b] -\shb{}_q^{(\textrm{\tiny{eq}})} \nonumber\\
\fl =4\pi\gamma_0\left[\int_0^{r_b}{\rho_{q0}(1+\Drho_0)\,\bar r^2\,\ln\rho\,\dd \bar r}-\ln\rhoav_q[r_b]\int_0^{r_b}{\rho_{q0}(1+\Drho_0)\,\bar r^2 \,\dd \bar r}\right],\label{sHBqq}
 \ea\ese
where we assume integration over complete time slices (no intersection with $t=\tbb( r)$ as in section 7.2) and we used (\ref{rhoHHKK}), (\ref{rhoKK}) and (\ref{Drho}). 

 We obtain  the asymptotic forms of $\rho,\,\rhoav_p[r_b],\,\rhoav_q[r_b]$ by substitution of (\ref{asrhok}) and (\ref{asaG}) in (\ref{rhoHHKK}), (\ref{rhoKK}), (\ref{Drho}), (\ref{rhoave2}) and (\ref{rhoaveq2}). Following the convergence of integrals described in Appendix B of \cite{RadAs}, we substitute these asymptotic forms together with (\ref{asrhok}) into the integrals (\ref{sHBpp})--(\ref{sHBqq}) and evaluate their asymptotic forms. After tedious algebraic manipulations we finally obtain the following results:
\begin{itemize}
\item Hyperbolic and elliptic models converging to Minkowski: both $\shb{}_p[r_b]$ and $\shb{}_q[r_b]$ diverge as $r_b\to\infty$.
\item Models converging to a FLRW state: the convergence of $\shb{}_p[r_b]$ and $\shb{}_q[r_b]$ strongly depends on the details of the convergence of $\rho$ and $\KK$, which is determined by the convergence of $\rho_{q0}$ and $\KK_{q0}$ (see \cite{RadAs}). Considering the case $\alpha=\beta=\nu$ (see Table 1) we obtain
\begin{itemize}
\item Hyperbolic models converging to open FLRW ($\nu>0$)
\bse\ba\fl \shb{}_p[r_b] -\shb{}_p^{(\textrm{\tiny{eq}})}\sim r_b^{2(1-\nu)},\qquad \shb{}_p[r_b]\,\,\hbox{converges for}\,\,\nu\geq 1,\\ \fl \shb{}_q[r_b] -\shb{}_q^{(\textrm{\tiny{eq}})}\sim r_b^{3-2\nu},\qquad \shb{}_q[r_b]\,\,\hbox{converges for}\,\,\nu\geq 3/2, \ea\ese
\item Hyperbolic models converging to Milne ($0<\nu\leq 3$)
\bse\ba\fl \shb{}_p[r_b] -\shb{}_p^{(\textrm{\tiny{eq}})}\sim r_b^{2-\nu/2},\qquad \shb{}_p[r_b]\,\,\hbox{diverges},\\ \fl \shb{}_q[r_b] -\shb{}_q^{(\textrm{\tiny{eq}})}\sim r_b,\qquad\qquad \shb{}_q[r_b]\,\,\hbox{diverges}, \ea\ese
\item Hyperbolic and elliptic models converging to Einstein de Sitter ($0<\nu\leq 2$)
\bse\ba\fl \shb{}_p[r_b] -\shb{}_p^{(\textrm{\tiny{eq}})}\sim r_b^{2-3\nu/2},\qquad \shb{}_p[r_b]\,\,\hbox{converges for}\,\,\nu\geq 4/3,\\ \fl \shb{}_q[r_b] -\shb{}_q^{(\textrm{\tiny{eq}})}\sim r_b^{3-2\nu},\qquad \shb{}_q[r_b]\,\,\hbox{converges for}\,\,\nu\geq 3/2, \ea\ese
\end{itemize}
\end{itemize}
Evidently, the exact value of $\nu$ that distinguishes convergence form divergence may change if we relax the condition $\nu=\alpha=\beta$ (see Table 1). However, for whatever values of these exponents allowed by regularity conditions, the following results emerge: 
\begin{itemize}
\item $\shb{}_p[r_b]$ and $\shb{}_q[r_b]$ diverge, and thus scale with volume ({\it i.e} they are ``extensive'' entropies) for models that radially converge to vacuum states (Milne and Minkowski). 
\item $\shb{}_p[r_b]$ and $\shb{}_q[r_b]$ converge (and thus do not scale with volume and are ``non--extensive'') when $\rho$ converges sufficiently fast to the density of a non--vacuum FLRW asymptotic state (open FLRW or Einstein--de Sitter).
\end{itemize}
The relation between the convergence of these entropies and the radial asymptotic behaviour of the density modes is evident: both entropies converge for models radially converging to non--vacuum FLRW for which both modes $\Jg$ and $\Jd$ vanish in the radial asymptotic limit (as all perturbations and fluctuations vanish in this limit for these models \cite{RadAs}). On the other hand, assuming (\ref{asrhok}) for models converging to Minkowski (see Table 1), we have  $\Drho_{q0}\sim -\alpha/3,\,\DKK_{q0}\sim-\beta/3$, thus (\ref{gmode})--(\ref{ivpconstr}), (\ref{asaG}) and (\ref{HTq})--(\ref{H32}) lead to:
\begin{itemize}
\item MD (matter dominated models, see Table 1): $\Omega_q\sim 1+O(\Omega_{q0}-1)\to 1$, and thus
\begin{equation}\fl \Jg \sim -\frac{\alpha-\frac{3}{2}\beta}{3-\alpha}\,O(\Omega_{q0}-1),\qquad \Jd\sim -\frac{\beta}{2(3-\alpha)}+O(\Omega_{q0}-1),\end{equation}
\item VD (vacuum or curvature dominated models, see Table 1): $\Omega_q\sim O(\Omega_{q0})\to 0$, and thus
\begin{equation}\fl \Jg \sim -\frac{\alpha-\frac{3}{2}\beta}{3(3-\alpha)}+O(\Omega_{q0}),\qquad \Jd\sim -\frac{\beta}{2(3-\alpha)}+O(\Omega_{q0}).\end{equation}
\end{itemize}
As a consequence, for $\alpha<3$ we have for MD models $\shb{}_p$ and $\shb{}_q$ diverging under conditions that are similar to those of early times for the general case: $\Jg\to 0$ and $\Jd$ finite so that $|\Jd|\gg |\Jg|$, while for VD models the entropies diverge under different asymptotic conditions: both modes tend to finite values with (in general) $|\Jg|> |\Jd|$ (notice that $\tbb'\to-\infty$ holds in all models radially asymptotic to Minkowski \cite{RadAs}, but $\tbb'$ is not the only factor involved in the decaying mode). In the case $\alpha=3$ (so that $\rho_{q0}\sim r^{-3}$, asymptotically Schwarzschild models) both entropies and both modes diverge for MD and VD models and thus further examination is required to verify which mode is dominant. 

\section{Summary and conclusion.} 

We have undertaken in this paper a comprehensive study of the application to generic LTB dust models of two different definitions of a gravitational entropy (section 2): the CET (Clifton, Ellis and Tavakol) proposal \cite{CET}, and two variants of the HB (Hosoya and Buchert) proposal: the original one (denoted by HBp) based on Buchert's average \cite{HB1,HB2,HB3} and one (denoted by HBq) constructed with a weighted  average (the q--average) and specially suited for LTB models \cite{part1}. In order to probe these entropy proposals on LTB models, we described (section 3) their dynamical and geometric properties by means of an initial value parametrization of their metric, together with a covariant representation of q--scalars, their fluctuations and perturbations \cite{part1,part2} expressed in terms of exact generalizations of the density growing and decaying modes of linear perturbation theory \cite{sussmodes}. We summarize below our main results:

\begin{description}
\item[Conditions for entropy growth.] The usage of q--scalars and their perturbations allowed for a unifying description of the necessary and sufficient conditions for non--negative entropy production for the three proposals described above (conditions (\ref{CETcond}), (\ref{HBcond}) and (\ref{HBcondq})):
\ba \fl \dot \shb{}_p \geq 0 \quad \Leftrightarrow\quad \left\langle\, (\rho-\rhoav_p)(\HH-\HHav_p)\,\right\rangle_p\leq 0,\qquad\hbox{HBp proposal},\nonumber\\
\fl \dot \shb{}_q \geq 0 \quad \Leftrightarrow\quad \left\langle\, (\rho-\rhoav_q)(\HH-\HHav_q)\,\right\rangle_q\leq 0,\qquad\hbox{HBq proposal},\nonumber\\
\fl \dot\sgr \geq 0 \quad \Leftrightarrow\quad (\rho-\rho_q)(\HH-\HH_q)\leq 0,\qquad\qquad\quad\hbox{CET proposal},\nonumber\ea
where we remark that the condition above for the CET proposal is a new result not obtained in the original CET paper \cite{CET}, while the conditions for the HBp and HBq proposals were previously known \cite{HB1,HB2,HB3,part1}. Entropy growth for the three proposals is directly related to a non--positive correlation of analogous (though strictly different) fluctuations of $\rho$ and $\HH$. We regard this finding as an appealing and important result. However, as discussed in section 5, there are subtle differences between these correlations:
\begin{itemize}
\item Entropy growth in the HBp and HBq proposals involves, strictly speaking, statistical correlations (statistical covariance moment) involving average functionals $\rhoav_p,\,\HHav_p$ and $\rhoav_q,\,\HHav_q$. 
\item Entropy growth in the CET proposal involves local correlations that are not statistical, as they involve the local functions (q--scalars) $\rho_q,\,\HH_q$ whose correspondence rule is the same as $\rhoav_q,\,\HHav_q$.  
\end{itemize}
As a consequence, the CET condition only needs to be evaluated locally at each point, whereas the HBp and HBq entropy production conditions are domain dependent, and thus are necessary and sufficient {\it only if} the integrals in the involved averages are evaluated for each given domain. Because of the integral non--local nature of these entropies, we can have entropy growth for a given domain $\DD[r_b]$, even if it decreases in local regions inside the domain (or in a smaller domain $\DD[r_c]\subset \DD[r_b]$ with $r_c\ll r_b$).  However, it is possible to obtain weaker (sufficient but not necessary) entropy production conditions for the HB entropies in terms of a uniform behavior of local fluctuations (see conditions (\ref{HBcond1})--(\ref{HBcond2}) and (\ref{HBcond3})). In particular, since $\rho_q,\,\HH_q$ coincide with the functionals $\rhoav_q,\,\HHav_q$ at the boundary of any domain, the entropy production condition for the CET entropy above is also an entropy production condition for the HBq entropy, but in the latter it is only sufficient and in the former it is necessary and sufficient.
\item[The CET entropy in the asymptotic evolution time ranges.] We proved analytically (section 6) that the CET entropy grows in all asymptotic evolution ranges in which the decaying mode is subdominant or is suppressed: asymptotic time range of expanding hyperbolic models, maximal expansion and collapse of elliptic models and near a simultaneous Big Bang (which follows by suppressing the decaying mode). The CET entropy decreases only in models (such as parabolic models) in which the growing mode is (artificially) suppressed and (for  general models) near the non--simultaneous Big Bang where the decaying is always dominant.  
\item[The HB entropies near singularities.] Because of their domain dependent nature the HBp/HBq entropies  exhibit a different behavior from that of the CET entropy near the non--simultaneous Big Bang and Big Crunch (see figure 2) in models with non-zero decaying mode: while the CET entropy diverges as dust layers reach these singularities, the HBp/HBq entropies are bounded in ``shrunk'' domains intersecting the singularities, though their time derivatives (entropy production) diverge for these domains (all this was proven in Appendix D.2). 
\item[Qualitative time evolution and numerical examples.] The asymptotic time behavior of the CET entropy production condition (section 7) yields sufficient information to put together a complete qualitative description of its full time evolution. This  evolution is depicted in Figures 1 for a typical dust layer. Considering dust layers in models for which the condition (\ref{CETcond}) for $\dot\sgr\geq 0$ holds (depicted by figures 1), and using the sufficient conditions (\ref{HBcond1})--(\ref{HBcond2}) and (\ref{HBcond3}), we can infer qualitatively the time evolution of both HB entropies in domains bounded by these dust layers in the non--asymptotic time range (see section 7). This evolution is depicted by figures 3 for a typical domain. We added (section 8) four numerical examples that fully corroborate the qualitative results on the CET entropy: a void model with suppressed decaying mode that fits supernovae and age constraints \cite{February:2009pv} (Figure 4), a similar void model but with non-zero decaying mode to compare with the previous example (Figure 5), and a ``spherical collapse model'' made of an elliptic collapsing region in a hyperbolic expanding exterior (Figure 6).
\item[Terminal entropy and net entropy gain.] The time asymptotic behaviour of the HBp/HBq entropies for ever-expanding models reveals (see proof in Appendix D.3) that there is a net entropy gain for the full time evolution of these models (Figures 3a and 3b), as the asymptotic value is a domain dependent ``equilibrium'' terminal value necessarily larger than the initial ``equilibrium'' value at the Big Bang.  There is also an analogous behaviour (a position dependent terminal equilibrium value) in the CET entropy applied to these models and locally evaluated along each local dust layer in the asymptotic time range (Figures 1a and 1b), though in this case the net entropy gain does not hold for the full time evolution because $\sgr\to\infty$ as $t\to\tbb$ (moreover, such entropy net gain occurs in the late time evolution: see Figures 1a and 1c, and if the decaying mode is suppressed: see Figures 1b and 1d). These entropy gains are a consistent result, since the CET and HBp/HBq entropy productions are positive as the evolution proceeds and the growing mode becomes dominant.  The HBp/HBq and CET entropies are more similar to each other in models with suppressed decaying mode (convergence results for these models were proven in Appendix D.4).
\item[Radial scaling of the CET entropy.] Since the CET entropy is defined by its entropy production law ($\dot\sgr$) through a Gibbs one--form, we solved the corresponding integrability condition to obtain its radial gradient $\sgr'$ (section 9). From this result we were able to provide analytic proof of the convergence of the CET entropy in the radial asymptotic range for all the different classes of radial asymptotic convergence of LTB models studied and classified in \cite{RadAs} (see Table 1). Since both $\dot\sgr$ and $\sgr'$ vanish as $r\to\infty$, the radial asymptotic ``equilibrium'' state must be characterized by a time independent constant $\sgr$ (as $\sgr$ is defined up to an additive constant) for all models. This result is consistent with the construction of the CET entropy with the Weyl and Bell--Robinson tensors, as LTB models converge asymptotically in the radial direction to spacetimes (FLRW or Minkowski) for which these tensors vanish.
\item[Radial scaling of the HB entropies.] Probing the radial asymptotic convergence of the HB entropies involved looking at these functionals for increasingly large domains in a given (complete) time slice. Using polynomial asymptotic trial functions derived in \cite{RadAs} we obtained the following results: the HBp/HBq entropies tend to an asymptotic time dependent ``equilibrium'' value in models radially converging to a FLRW background, but {\it only} if $\rho$ and $\HH$ converge sufficiently fast to their asymptotic FLRW values. For a slow convergence to FLRW and for models converging to a vacuum state (Minkowski or Milne), these entropies diverge in the asymptotic radial range.

\end{description} 

\subsection{The CET gravitational entropy and isotropic cosmological singularities.} 

We have shown from the asymptotic and qualitative study of the CET entropy that: 
\begin{itemize}
\item $\sgr$ decreases in time ranges in which the decaying mode $\Jd$ is dominant, either when the growing mode $\Jg$ is suppressed (section 6.3) or if $\Jg\ne 0$ as dust layers emerge from a non--simultaneous Big Bang. 
\item $\sgr$ increases near the Big Bang singularity and throughout the full time evolution {\it only} in models examined in section 6.2 in which the decaying mode is totally suppressed: $\Jd=0$  (this is also true for the HBp/HBq entropies). 
\end{itemize} 
However, as shown by Wainwright and Andrews \cite{WainAndr}, LTB models with a suppressed decaying mode emerge from an isotropic Big Bang and converge in earlier times to a spatially flat EdS model. This suggests an important theoretical connection between the early time behavior of the CET entropy and basic geometric features of the initial singularity and early time evolution of the models. In particular, the comprehensive studies by Goode and Wainwright \cite{GooWain} and Lim {\it et al} \cite{Limetal} may further suggest that an ever increasing CET entropy may also be a characteristic robust property of generic inhomogeneous perfect fluid models ($\Lambda\ne 0$ is assumed in \cite{Limetal}) admitting an isotropic initial singularity associated with an early times convergence to a spatially flat FLRW model (EdS model). Lim {\it et al} characterize this class of models by only three free parameters, which for the case of pure dust sources ($p=\Lambda=0$) reduce to a single free function. Evidently, LTB models with $\Jd=0$ are the spherically symmetric dust sub--case (with $\Lambda=0$) of this class of generic models defined by Lim {\it et al}, as they are fully specified by a single free function (see \cite{sussmodes}) and (as shown by \cite{WainAndr}) they satisfy the asymptotic conditions that \cite{GooWain} and \cite{Limetal} use to define the isotropic Big Bang and early time EdS behavior. Since a full detailed comparison with the formalism of \cite{GooWain} and \cite{Limetal} is outside the scope of this paper, we examine only their asymptotic condition $\hat\Omega=1$ near the initial singularity, where $\hat\Omega$ is given by: 
\begin{equation} \hat \Omega \equiv \frac{8\pi\,\rho}{3\HH^2} = \Omega_q \frac{1+\Drho_q}{(1+\Dh_q)^2},\qquad \Omega_q = \frac{8\pi\,\rho_q}{3\HH_q^2},\end{equation}
and must not be confused with the q--scalar $\Omega_q\ne \hat\Omega$ defined by (\ref{OmDOm}) (Lim {\it et al} denote $\hat\Omega$ by ``$\Omega$'' and this can be confusing with our notation). Since $\Omega_q\to 1$ holds for all LTB models as $t\to\tbb$ (irrespective of whether the decaying mode is suppressed or not), then the early time EdS behavior $\hat\Omega\to \Omega_q\to 1$ as $t\to \tbbo$ occurs only if $\Jd=0$, as for these models we have $\Drho,\,\Dh\to 0$ as $t\to\tbbo$ (see equation (\ref{DrhoDh0})). However, LTB models with a nonzero decaying mode ($\Jd\ne 0$) do not comply with the conditions of \cite{GooWain} and \cite{Limetal} for an isotropic Big Bang and EdS behavior, as we have for these models $\Drho_q\to -1$ and $\Dh_q\to-1/2$ (see (\ref{JgJd1}) and (\ref{deltasbb})), so that $\hat\Omega\to 0$ holds as $t\to\tbb$. 

Lim {\it et al} proved that the isotropic Big Bang and associated EdS behavior are robust geometric features preserved by a change of frame (``time gauge'' in the sense of a 4--velocity boost \cite{coley}), hence the early time increasing behavior of $\sgr$ that we have proved for the CET entropy in models with $\Jd=0$ is also robust in this sense.  However, the CET and HBp/q entropies in the late time evolution and/or models with $\Jd\ne 0$ are likely to be affected by such a change of frame, since the entropy growth conditions (\ref{CETcond}), (\ref{HBcond}) and (\ref{HBcondq}) also depend on fluctuations of the Hubble scalar $\HH$ related to the eigenvalue of the shear tensor (see (\ref{sigEC})--(\ref{SigPsi2})), which is sensitive to the choice of 4--velocity. Looking at this issue in detail is beyond the scope of the present paper. 

It is also important to emphasize that an isotropic singularity is not a necessary condition for a non--negative CET entropy production, as we have $\dot\sgr\to\infty$ as dust layers approach the collapse singularity, which is not an isotropic singularity in the sense of \cite{GooWain,Limetal} (notice that $\hat\Omega\to 0$ as $t\to\tcoll$ in elliptic models). However, entropy production  from the the HBp/q entropies become negative as $t\to\tcoll$.      

\subsection{Gravitational entropy vs cosmological ``homogenization''.}

Our results (summarized before) are in excellent agreement with those of a recent numerical study by Bolejko and Stoeger \cite{bolstoeg}, who considered various entropy proposals (including the old ``arrow of time'' notion) in spherically symmetric models endowed with a rather general matter content: general perfect fluids and anisotropic fluids with non-zero viscosity. These authors showed that there is always a period in the evolution of their models in which entropy (in its various definitions) decreases, an effect they associate with a process in which the Universe "homogenizes" when the decaying mode is present and dominant. Evidently, this result is identical to our analytical findings on the early time behaviour of the CET and HBp/HBq entropies in models with a non-zero decaying mode (see section 7 and Figures 1a, 1c, 3a, 3c, 5b and 6). Furthermore, they show that, after this ``homogenization'' stage, their models evolve with increasing entropy towards an asymptotic terminal inhomogeneous state (they did not examine collapsing configurations). Again, this numerical result is identical to our analytic and qualitative result that the initially decreasing entropies  begin to grow until reaching a (position or domain dependent) terminal profile in the asymptotic time range of hyperbolic models (see section 7, Appendix D.3 and Figures 1a, 1b, 3a, 3b). While Bolejko and Stoeger examined more general spherically symmetric models, we were only concerned with LTB dust models. Hence, they had to rely on a numerical treatment in which the identification of the (necessarily coupled) growing or decaying density modes of fully non--linear sources is practically impossible, whereas we were able to undertake a fully analytic treatment of the CET and the two HB entropies using the analytic (exact) non--linear forms for these coupled modes obtained for LTB models in \cite{sussmodes}. 

As a further note: Bolejko and Stoeger identify the ``homogenization'' phase of their models with a sort of Einstein--de Sitter unstable saddle point: this result was obtained rigorously for LTB models in the dynamical systems study undertaken in \cite{sussmodes}.               

\subsection{The gravitational entropy and the cosmological constant.}

While we have only considered gravitational entropy for LTB models with $\Lambda=0$, important qualitative information on the entropy growth, at least for the CET entropy, follows from previous work on the dynamics of LTB models with $\Lambda>0$ \cite{sussDS2}. Since the invariant scalar $\Psi_2$ (and thus $\rhogr$) is the same when we consider a $\Lambda>0$ term, the condition for the CET entropy growth is also given by (\ref{CETcond}) in this case. Hence, figures 2, 3, 4, 5 and 8 of \cite{sussDS2} provide numerical examples in which $\Drho\Dh<0$, and thus $\dot\sgr>0$, hold in the asymptotic time range of ever--expanding  LTB models with $\Lambda>0$. However, a proper examination of the gravitational entropies for generic LTB models with $\Lambda> 0$ requires a fully separate study, which is relevant given the fact that these models are an inhomogeneous generalization of the $\Lambda$CDM model.

\subsection{The gravitational entropy and the need to suppress the decaying mode.}              

The full time non--negative entropy production when the decaying mode is totally suppressed ($\Jd=0$) seem to lend further weight to the preference of using models with this feature in cosmological and astrophysical applications, in particular in the effort to fit cosmological observations without assuming the existence of dark energy \cite{Zibin:2011ma,Bull:2011wi} (though see \cite{CBK,clarkreg} for a critical approach to this line of thought). However, we believe that our results do not justify the total suppression of the decaying mode, which is an excessively extreme and not strictly necessary option, as the early times negative entropy production associated with a decaying mode can be regarded merely as a signal that LTB models are no longer viable for these radiation dominated cosmic times. In fact, a non--relativistic dust source with zero pressure gradients is not expected to be a physically plausible matter model near any singularity, not even an isotropic singularity (in the sense of \cite{GooWain,Limetal}) associated with a positive entropy production like the simultaneous Big Bang of models with $\Jd=0$.  While CMB constraints and compatibility with the inflationary paradigm require nearly homogeneous conditions for sufficiently early times at the onset of (non-relativistic) matter dominated era, this requirement can be met by LTB models (such as our second numerical example in section 8) with a decaying mode that has become sufficiently subdominant (though not totally suppressed) at the required times, discarding the evolution of the model for previous times when radiation is dominant.

\subsection{Stability and extensivity.}

Considering the relation between the type of the extrema of the entropy ({\it i.e.} concavity of its time profile) and the stability of equilibrium states, we can associate the perpetual expansion of hyperbolic models with a time asymptotic entropy maximum for all proposals (see the convex time profile of the curves of Figures 1a, 1b, 3a and 3b for large times), which corresponds to a stable asymptotic terminal equilibrium state. On the other hand, the concave time profiles of the three entropies, either in regions where decaying modes are dominant (early times in Figures 1a, 1c, 2a and 2c), or in elliptic models (Figures 1c, 1d, 3c, 3d), suggest that decaying modes and collapsing configurations should be characterized by unstable equilibrium states of the gravitational entropy. This unstable equilibrium state is reminiscent of the unstable equilibria that characterize the Boltzmann--Gibbs entropy in non--collisional Newtonian systems subjected to relaxation processes and Antonov's instability \cite{Newtonian}. However, the gravitational entropy is a different concept from the Boltzmann--Gibbs entropy and these Newtonian systems evolve towards terminal stationary states,  while a stationary state in elliptic LTB models only arises at the instant of maximal expansion. Nevertheless, despite these differences, this correspondence between the unstable equilibria of self gravitating systems in all these entropies is worth exploring, specially by looking in future research at the gravitational entropy proposals for general relativistic stationary (or asymptotically stationary) systems.

Another point worth commenting upon is the issue of the ``extensive'' or ``non--extensive'' nature of the CET and the two HB entropies. If we define an extensive quantity as scaling with volume, then the CET entropy is clearly non--extensive for all LTB models, while the HB entropies are only non--extensive for models converging sufficiently fast to a FLRW background. The HB entropies for slow convergence to FLRW or convergence to vacuum states are then extensive. On the other hand, if we regard large decaying modes as unphysical, then the extensive nature of the two HB entropies in LTB models converging to vacuum states seems to be consistent with the viability of a non--extensive entropy, as some of these models exhibit large decaying modes in the asymptotic radial range (though further study is needed to find out which one of the decaying or growing mode is dominant). These facts point out to a possible theoretical connection with studies of thermodynamical properties of Newtonian self--gravitating systems, which show that the long range nature of gravity blurs the simple distinction between intensive and extensive thermodynamical variables and may lead to non--extensive energy and entropy \cite{Newtonian}.   A possible theoretical connection to Tsallis' non--extensive entropy \cite{Tsallis} may also be worth exploring in future research.    

\subsection{Final comments.}  

We would like to highlight the utility of describing LTB models with the q--scalar representation, as the original LTB variables would not have allowed us to link the CET entropy production to fluctuations of $\rho$ and $\HH$, which  resemble the fluctuations that result from the the two HB entropies, and thus provide the key theoretical unifying connection among all these entropies. This fact also provides a strong motivation to study possible extensions of the q-scalars to spacetimes more general than LTB (q--scalars have been used for the study of LTB models with $\Lambda>0$ in \cite{sussDS2}), and even non-spherical spacetimes, in order to examine general theoretical properties of the inhomogeneous gravitational field (including the study of the CET and HB entropies in less symmetrical contexts). We are currently elaborating a follow up paper to deal with the gravitational entropies in the case $\Lambda>0$ and for the non--spherical Szekeres dust models, as formal results valid for LTB models can be easily generalized for the latter models along the lines of \cite{sussbol}. Spacetimes with perfect fluid and dissipative sources, as well as the ``wave--like'' Petrov type N spacetimes examined by \cite{CET}, are also important candidates for further investigation.   

\section*{Acknowledgments:} RAS acknowledges support from Mellon Foundation during a visit to Rhodes University, when this work was initiated. RAS also acknowledges financial support from grant CONACYT 132132. The authors thank Timothy Clifton, George Ellis, Thomas Buchert and Krzysztof Bolejko for useful comments and discussions.

\begin{appendix}

\section{LTB models in their standard variables.}

LTB dust models are usually given by the following traditional metric: 
\begin{equation} \dd s^2 =-\dd t^2+\frac{R'{}^2}{1+2E}\,\dd r^2 +R^2\left(\dd\vartheta^2+\sin^2\vartheta\,\dd\varphi^2\right),\label{ltb1}\end{equation}
where $R=R(t,r),\,E=E( r),\,R'=\partial R/\partial r$  and $R$ satisfies the Friedman--like equation
\begin{equation}\dot R^2 = \frac{2M}{R}+2E,\label{Friedeq1}\end{equation}
with $M=M_q( r)$ (the quasi--local mass--energy functional in (\ref{Mq}) but treated as a function). The Friedman--like LTB metric (\ref{ltb2}) the Friedman--like equation (\ref{HHq}) follow from (\ref{ltb1}) and (\ref{Friedeq1}) by selecting the radial coordinate such that $R_0=R(t_0,r)=r$ for an arbitrary fiducial hypersurface $t=t_0$ and defining
\begin{equation} a\equiv \frac{R}{r},\qquad \Gamma =\frac{rR'}{R}=1+\frac{ra'}{a},\end{equation}
so that $a_0=\Gamma_0=1$. The relation between the free functions $M$ and $E$ in (\ref{ltb1})--(\ref{Friedeq1}) and the basic initial value q--scalars $\rho_{q0},\,\KK_{q0},\,\HH_{q0},\,\Omega_{q0}$ is given by
\footnote{Under this initial value parametrization the ``Big Bang time'' $\tbb$ follows as a function of any two basic initial q--scalars. See equation (\ref{tbb}). }
\ba \fl \frac{8\pi}{3}\rho_{q0}=\frac{2M}{r^3}=\Omega_{q0}\HH_{q0}^2,\qquad \KK_{q0}=-\frac{2E}{r^2}=(\Omega_{q0}-1)\HH_{q0}^2,\label{ivf1}\\
\fl \HH_{q0}=\left[(8\pi/3)\rho_{q0}-\KK_{q0}\right]^{1/2}=\frac{[2M+2Er]^{1/2}}{r^{3/2}},\quad \Omega_{q0}=\frac{8\pi\rho_{q0}}{3\HH_{q0}^2}=\frac{M}{M+Er},\label{ivf2}\ea
with their initial value perturbations $\Da_{q0}$ obtained from (\ref{Dadef}) with $\Gamma=1$. The main covariant scalars in (\ref{rhoHHKK}) and their corresponding q--scalars take the following form in terms of the standard variables:
\ba\fl  4\pi\rho =\frac{M'}{R^2R'},\qquad  \HH = \frac{\partial_t(R^2 R')}{3R^2 R'},\qquad
\KK = -\frac{4(ER)'}{R^2 R'},\label{basic}\\
\fl \frac{4\pi}{3}\rho_q =\frac{M}{R^3},\qquad  \HH_q = \frac{\dot R}{R},\qquad
\KK_q = -\frac{2E}{R^2}.\label{qbasic}\ea
The eigenvalues of the shear, electric Weyl and Weyl tensors in the standard variables are
\begin{equation} \Sigma = -\frac{\partial_t(R'/R)}{3\,R'/R},\qquad \Psi_2=\frac{M}{R^3}-\frac{4\pi}{3}\rho.\label{SigPsi22}\end{equation}
It is straightforward to show that these forms are identical to those in (\ref{SigPsi2}).

\section{Analytic solutions.}

The analytic solutions of the Friedman equation (\ref{HHq}) (equivalent to (\ref{Friedeq1}))
\begin{equation} \HH_q =\frac{\dot a}{a}=\HH_{q0}\left[\frac{\Omega_{q0}}{a^3}-\frac{\Omega_{q0}-1}{a^2}\right]^{1/2},\end{equation}
can be given in terms of $\Omega_q$ for initial conditions $\HH_{q0},\,\Omega_{q0}$ as
\footnote{The ``parabolic'' case follows from the elliptic and hyperbolic cases as the limit $\Omega_{q0}\to 1$ (or $E=0$ in (\ref{Friedeq1})). For these models $\HH_q(t-\tbb)=2/3$ holds exactly, hence $\Jg=0$ even if $\Dig\ne 0$ (see \cite{sussmodes}).} 
\ba
\fl {\underline{\hbox{hyperbolic models:}}}\,\,0<\Omega_q<1\,\,(\KK_q<0\,\,\,\hbox{or}\,\, E>0),\qquad , \nonumber\\  
 t-\tbb = \frac{Y_q(\Omega_q)}{\HH_q},\label{hypsol}\\
 \fl {\underline{\hbox{elliptic models:}}}\,\,\Omega_q>1\,\,(\KK_q>0\,\,\,\hbox{or}\,\, -1<E<0),\qquad ,\nonumber\\
\fl t-\tbb= \left\{ \begin{array}{l}
 Y_q(\Omega_q)/\HH_{q},\qquad\qquad\qquad  
 {\hbox{expanding phase}}\quad \HH_q>0,\\ 
 2\pi\beta_q - Y_q(\Omega_q)/\HH_q, \qquad 
 {\hbox{collapsing phase}}\quad \HH_q<0,\\ 
 \end{array} \right.\label{ellsol}\ea
with the functions $\beta_{q}$ and $Y_q$ given by
\ba  
\beta_q = \frac{4\pi\rho_q}{3|\KK_q|^{3/2}}=\frac{\Omega_q}{2|1-\Omega_q|^{3/2}\HH_q}=\beta_{q0}\qquad\left(\Rightarrow \dot\beta_q=0\right),\label{beta}\\
 Y_q(\Omega_q) = \frac{\epsilon}{|1-\Omega_q|}\left[1-\frac{\Omega_q}{2|1-\Omega_q|^{1/2}}\ACal\left(\frac{2}{\Omega_q}-1\right)\right], \label{Y}
\ea
where $\epsilon =1,\, \ACal=$ arccosh correspond to the hyperbolic case and $\epsilon =-1,\, \ACal=$ arccos to the elliptic case. Notice that these solutions can be given in terms of the scale factor $a$ by substituting the form of $\Omega_q$ in (\ref{OmDOm}) into the right hand sides of (\ref{hypsol}) and (\ref{ellsol}) (see Appendix A2 of \cite{sussBR}).  
 
The Big Bang time $\tbb=\tbb(r )$ and its gradient are expressible in terms of primary initial value functions and perturbations:
\begin{equation} \tbb=t_0-\frac{Y_q(\Omega_{q0})}{\HH_{q0}},\qquad r\tbb' = (1+\Drho_{q0})\Did,\label{tbb}\end{equation}
where we substituted $t=t_0$ in (\ref{hypsol}) and (\ref{ellsol}), while the form of $\tbb'$ follows from (\ref{Did}). Besides $\tbb$, elliptic models have the following characteristic times:
\ba 
\fl \hbox{elliptic expanding:}\quad 0< t-\tbb \leq \tmax-\tbb=\pi\beta_{q0},\label{tmax}\\
\fl \hbox{elliptic collapsing:}\quad  \tmax-\tbb < \tau_q<\tcoll-\tbb=2\pi\beta_{q0},\label{tcoll}
\ea
where $t=\tmax$ and $t=\tcoll$ mark the times of maximal expansion ($\HH_q=0$) and the collapse singularity (``Big Crunch'' $\HH_q\to-\infty$). 

The perturbations and fluctuations become fully determined once we compute $\Jg$ and $\Jd$. For this purpose, we need the following expressions:   
\ba
\fl \HH_q(t-\tbb)=Y_q,\qquad \hbox{(hyperbolic \& elliptic expanding)},
\label{HTq}\\
\fl \HH_q(t-\tbb)=Y_q-\frac{\pi\Omega_q}{(\Omega_q-1)^{3/2}},\quad\hbox{(elliptic collapsing)}, \label{HTqc}\ea
\begin{equation}\fl \HH_q =\pm\HH_{q0}\frac{\Omega_q}{\Omega_{q0}}\,\left[\frac{1-\Omega_{q0}}{1-\Omega_q}\right]^{3/2},\label{H32}\end{equation}
where $Y_q=Y_q(\Omega_q)$ is given by (\ref{Y}), $\epsilon = 1,-1$ correspond to hyperbolic and elliptic cases and we can use the scaling law (\ref{OmDOm}) to express $\HH_q(t-\tbb)$ above in terms of the scale factor $a$ and initial value functions.

\section{Evolution equations.}

While the dynamics of LTB models is fully determined analytically, they can also be studied numerically by solving  the evolution equations for the q--scalars and their perturbations given by (19a)--(19d) or (21a)--(21d) of \cite{part2}.  However, these evolution equations can be problematic for collapsing configurations, since $\Dh_q,\,\Omega_q$ and $\DOm$ diverge as $\HH_q\to 0$ (when $t\to\tmax$), and thus the expanding ($\HH_q>0$) and collapsing ($\HH_q<0$) stages of elliptic models must be treated separately. For a unified numerical treatment of collapsing configurations (as the third example of section 8) we used instead the following evolution equations for the q--scalars and their fluctuations $\Del_q(\rho),\,\Del_q(\HH)$ which are bounded at $t=\tmax$:
\ba \dot \rho_q &=& -3 \rho_q\HH_q,\label{EVq21}\\
\dot \HH_q &=& -\HH_q^2-\frac{4\pi}{3}\rho_q, \label{EVq22}\\
\left[\Del_q(\rho)\right]\dot{} &=& -3[\rho+\Del_q(\rho)]\,\Del_q(\HH)-3\HH_q\Del_q(\rho),\label{EVq23}\\
\left[\Del_q(\HH)\right]\dot{} &=& -\frac{4\pi}{3}\Del_q(\rho) - [2\HH_q+3\Del_q(\HH_q)]\Del(\HH_q).\label{EVq24}\ea

\section{Formal results on the HB entropies.}

\subsection{Non--negativity of the HB functionals.}

The HB entropy functional (\ref{sHB}) is strictly non--negative for $\rho\geq 0$ and any associated scalar average of the form (\ref{rhoav}) (which includes the quasi--local average (\ref{sHBq})). The proof (private communication from T. Buchert) follows by remarking that the inequality 
\begin{equation}\ln x \leq x-1 \qquad x = \frac{\rhoav_\DD}{\rho}\end{equation}
holds for all non--negative $x$. Multiplying both sides by $-\rho\leq 0$ and using the property $-\ln x=\ln x^{-1}$ leads to the desired result:
\ba \fl \rho\,\ln \left[\frac{\rho}{\rhoav_\DD}\right]\geq \rho-\rhoav_\DD\quad \Rightarrow\quad \shb =\left\langle \rho\,\ln \left[\frac{\rho}{\rhoav_\DD}\right]\right\rangle_\DD \VV_\DD\geq 0,\ea
since $\langle\, \rho-\rhoav_\DD\,\rangle_\DD =0$ holds for any scalar average (\ref{rhoav}).       

\subsection{Proof of the limits (\ref{shblim}) and (\ref{dotshblim}).}

We elaborate the convergence proofs only for $\shb{}_q$, as the proofs for $\shb{}_p$ are analogous. We look first at the case of slices intersecting the Big Bang, considering domains shown in Figure 2a in a slice $t=t_s\approx t_{(-)}$, such that $r_b>r_s$ (with $r_b\approx r_s$), where $t_s=\tbb(r_s)$, hence $t_s-\tbb( r)\approx -\tbb'{}_s(r-r_s)$ holds with $\tbb'{}_s=\tbb'(r_s)<0$ (from demanding absence of shell crossings). Since $\rho\to\infty$ as $r\to r_s$, the  convergence test of (\ref{sHBq}) follows by rewriting this integral as the limit
\ba \fl  \shb{}_q[r_b] -\shb{}_q^{(\textrm{\tiny{eq}})}=\gamma_0 \lim_{\epsilon\to 0}\int_{r_s+\epsilon}^{r_b}{M'_q(\bar r) \ln\left[\frac{\rho}{\rhoav_q[r_b]}\right]\dd\bar r},\qquad M'_q=4\pi\rho_{q0}(1+\Drho_{q0})r^2,\nonumber\\
\label{sHBq2}\ea
where $\epsilon>0$. Since $a\ll 1$ holds for the domains we are interested in, we use (\ref{deltasbb}) and expand the solutions (\ref{hypsol}) and (\ref{ellsol}) in this limit \cite{sussBR} to obtain at first order in $r-r_s$ 
\ba\fl a^{3/2}(t_s,r)\approx \left[\frac{3}{2}\sqrt{\Omega_{q0}}\HH_{q0}|\tbb'|\right]_s(r-r_s),\qquad 1+\Drho_{q}(t_s,r) \approx \frac{3(1+\Drho_{q0s})}{2r_s}(r-r_s),\nonumber\\ \label{approx1}\ea
where the subscript ${}_s$ will denote henceforth evaluation at $r=r_s$. While $\rho$ along $t=t_s$ can be directly computed by inserting (\ref{approx1}) in (\ref{rhoHHKK}) and (\ref{rhoKK}), the density average in this slice is no longer given by (\ref{rhoaveq2}), but by $\rhoav_q[r_b]=(M_{qb}-M_{qs})/\VV_q[r_b]$, with $M_{qb}=M_q(r_b)$ and $M_{qs}=M_q(r_s)$. Applying (\ref{approx1}) to the appropriate forms for $\rho$ and $\rhoav_a[r_b]$ we obtain 
\ba\fl \rho\approx \frac{M'_{qs}}{6\pi\,\Omega_{q0s}\HH_{q0s}^2|\tbb'{}_s|^2\,r_s(r-r_s)},\quad \rhoav_q[r_b]\approx \frac{M'_{qs}}{3\pi\, \Omega_{q0s}\HH_{q0s}^2 |\tbb'{}_s|^2\,r_s(r_b-r_s)},\nonumber\\\label{approx2}\ea
where we used $M_{qb}-M_{qs}\approx M'_{qs}(r_b-r_s)$. By inserting these forms into (\ref{sHBq2}) and taking the limit as $\epsilon\to 0$ we obtain:
\ba  \shb{}_q[r_b] -\shb{}_q^{(\textrm{\tiny{eq}})}= (1-\ln 2)\,M'_{qs}(r_b-r_s)>0,\label{approx3}
\ea
which implies the convergence of the integral (\ref{sHBq2}) for the domains under consideration. The second limit in (\ref{shblim}) follows readily as $r_s\to r_b$. For the HBp entropy (first limit in (\ref{shblim})) we obtain the same result as (\ref{approx3}), but with $M'_{qs}$ replaced by $M'_{ps}$. The collapse case yields the same result as (\ref{approx3}) for the HBp and HBq entropies, since for slices intersecting $t=\tcoll$ we have $a^{3/2},\,\rho$ and $\rhoav_q[r_b]$ taking the same forms (\ref{approx1}), but proportional to $\tcoll-t_s\approx \tcoll'(r_s)(r-r_s)$, hence the ratio $\rho/\rhoav_q$ has the same form as that obtained with (\ref{approx2}). 

In order to verify the convergence of $\dot\shb{}_q[r_b]$ in a slice $t=t_s$ intersecting the Big Bang, we re--write (\ref{Sdot1q}) as
\begin{equation}  \dot\shb{}_q[r_b]=\gamma_0\left[\lim_{\epsilon \to 0}\int_{r_s+\epsilon}^{r_b}{M'_q\,\HH\,\dd\bar r}-\HHav_q[r_b]\,(M_{qb}-M_{qs})\right],\label{dotSHB2}\end{equation}
where we used the identities $\rho \VV'_q =M'_q$ and $\rhoav_q[r_b]\,\VV_q[r_b]=M_{qb}-M_{qs}$ and we are taking into consideration that $\HH\to\infty$ as $r\to r_s$. Considering that $M'_q\approx M'_{qs}+M''_{qs}(r-r_s)$ and expanding $\HH$ and $\HHav_q[r_b]=\dot a(t_s,r_b)/a(t_s,r_b)$ along $t=t_s$ at first order in $r-r_s$ we obtain
\begin{equation} \HH\approx \frac{1}{3\,|\tbb'{}_s|(r-r_s)},\qquad  \HHav_q[r_b]\approx\frac{2}{3\,|\tbb'{}_s|(r_b-r_s)}, \label{approx4}\end{equation}
all of which inserted into (\ref{dotSHB2}) yields:
\begin{equation} \dot\shb{}_q[r_b]\approx \frac{\gamma_0 M'_{qs}}{3\,|\tbb'{}_s|}\left[\,\lim_{\epsilon \to 0}\int_{r_s+\epsilon}^{r_b}{\frac{\dd \bar r}{\bar r-r_s}}-2\right]\to \infty,\label{approx5}\end{equation}
where we used $M_{qb}-M_{qs}\approx M'_{qs}(r_b-r_s)$. We obtain the same result for $\dot\shb{}_p[r_b]$, with $M'_{qs}$ replaced by $M'_{ps}$. Since we keep $r_b>r_s$ fixed, the lack of convergence of these integrals implies the limits (\ref{dotshblim}).  

Near the collapsing singularity we obtain similar forms as in (\ref{approx4}) with $|\tbb'{}_s|$ replaced by $|\tcoll'{}_s|$, but now we have $\HH<0$ and $\HHav_a[r_b]<0$. Hence the right hand side of (\ref{approx5}) has the opposite sign, leading to $\dot\shb{}_q[r_b]\to -\infty$ and $\dot\shb{}_p[r_b]\to -\infty$ as $r\to r_s$, and thus the limits (\ref{dotshblimc}) follow.   

\subsection{The HB entropies in the asymptotic time range of hyperbolic models.}

To examine the asymptotic time range $t\gg t_0$ of hyperbolic models we expand (\ref{hypsol}) for $a\gg 1$ (or $\Omega_q\ll 1$) and use (\ref{JgJd2}) to obtain 
\begin{equation} a \approx |\KK_{q0}|^{1/2}\Delta,\qquad 1+\Drho\approx \frac{1}{1-\Dig}=\frac{1+\Drho_{q0}}{1+\frac{3}{2}\DKK_{q0}},\end{equation}
where $\Delta=t-t_0$. This yields 
\begin{equation}\fl  \rho\approx \frac{\rho_{q0}(1+\Drho_{q0})}{\Delta^3\,|\KK_{q0}|^{3/2}\left(1+\frac{3}{2}\DKK_{q0}\right) },\quad \rhoav_q[r_b]\approx \frac{\int_0^{r_b}{\rho_{q0}(1+\Drho_{q0}) \bar r^2\dd\bar r}}{\Delta^3\,\int_0^{r_b}{|\KK_{q0}|^{3/2}\left(1+\frac{3}{2}\DKK_{q0}\right)\bar r^2\dd\bar r} }, \end{equation}
which inserted into (\ref{sHBq2}) (with $r_s=0$) leads to the following asymptotic terminal expression as $t\to\infty$: 
\ba \fl \shb{}_q(r_b) -\shb{}_q^{(\textrm{\tiny{eq}})}\approx\nonumber\\
\fl \int_0^{r_b}{\rho_{q0}(1+\Drho_{q0})\,\bar r^2\,\ln\left[\frac{\rho_{q0}(1+\Drho_{q0})}{\langle\rho_{q0}(1+\Drho_{q0})\rangle_q[r_b]}\,\frac{\langle |\KK_{q0}|^{3/2}(1+\frac{3}{2}\DKK_{q0})\rangle_q[r_b]}{|\KK_{q0}|^{3/2}(1+\frac{3}{2}\DKK_{q0})}\right]\dd\bar r},
\ea
whose right hand side is obviously not zero, hence it must be positive (see proof of non--negativity in Appendix D.1). Therefore, the terminal value of $\shb{}_q[r_b]$ for all domains with $r_b>0$ is necessarily larger that the equilibrium initial value (either at $t=t_{(-)}$ when the decaying mode is non-zero or at $t=\tbbo$ when this mode is suppressed). The same results hold for $\shb{}_p[r_b]$. 

\subsection{Models with a suppressed decaying mode.}

For these models we evaluate the integral in (\ref{sHBq}) for slices $t\approx \tbbo$. Expanding (\ref{hypsol}) and (\ref{ellsol}) with $\tbb=\tbbo$ for $a\ll 1$ and using (\ref{DrhoDh0}) we obtain
\begin{equation}\fl a^{3/2}\approx \frac{3}{2}\sqrt{\Omega_{q0}}\HH_{q0}\Delta,\qquad 1+\Drho_{q} \approx 1-\frac{2(\Omega_{q0}-1)\Dig}{5\Omega_{q0}}\,a, \label{approx6}\end{equation}
where now  $\Delta\equiv t-\tbbo$. The density and its q--average take the form
\ba \fl \rho \approx \frac{4}{9\Delta^2}\left[1-\frac{18^{1/3}\Dig (\Omega_{q0}-1)(\HH_{q0})^{2/3}}{5(\Omega_{q0})^{2/3}\Delta^{2/3}}\right],\quad 
\rhoav_q[r_b]\approx \frac{4}{9\Delta^2}\left[1-\frac{M'_{qb}}{5M_{qb}\Delta^{2/3}}\right],\nonumber\\\ea 
so that $\rho( r)/\rho_q[r_b]\approx 1+F(r,r_b)\Delta^{2/3}$, where $F(r,r_b)$ follows from the initial value functions $\Omega_{q0},\,\HH_{q0},\,\Dig$ and the domain dependent constants $M_{qb},\,M'_{qb}=M'_q(r_b)$. Since $\ln[\rho( r)/\rhoav_q[r_b]]\approx F(r,r_b)\Delta^{2/3}$, the HBq entropy becomes
\begin{equation}\shb{}_q(r_b) -\shb{}_q^{(\textrm{\tiny{eq}})}\approx \frac{\Delta}{4\pi}\int_0^{r_b}{M'_q(\bar r) F(\bar r,r_b)\dd \bar r},\end{equation}
and thus the right hand side vanishes for all domains in the limit $\Delta\to 0$. Regarding the behaviour of $\dot\shb{}_q[r_b]$, we use the form (\ref{dotSHB2}) but for the whole integration range $0\leq r \leq r_b$, hence $M_{qs}=0$. Since $\HH$ and $\HHav_q[r_s]$ are both proportional to $1/\Delta$ for slices $t\approx \tbbo$, then $\dot\shb{}_q[r_b]\to\infty$ as $t-\tbbo$. The same result follows readily for $\dot\shb{}_p[r_b]$.      

\end{appendix}

\end{document}